\documentclass[a4paper,11pt]{article}%
\usepackage{geometry}
\usepackage{amsmath}
\usepackage{amsfonts}
\usepackage{amssymb}
\usepackage{graphicx}
\usepackage{indentfirst}
\usepackage{bbold}
\usepackage[small,bf]{caption}
\usepackage{slashed}
\usepackage{enumerate}
\providecommand{\U}[1]{\protect\rule{.1in}{.1in}}

\begin{document}

\date{}
\title{\textbf{Properties of the Faddeev-Popov operator in the Landau gauge, matter confinement and soft BRST breaking}}
\author{\textbf{M.~A.~L.~Capri}\thanks{caprimarcio@gmail.com}\,\,,
\textbf{M.~S.~Guimaraes}\thanks{msguimaraes@uerj.br}\,\,,
\textbf{I.~F.~Justo }\thanks{igorfjusto@gmail.com}\,\,,\\
\textbf{L.~F.~Palhares}\thanks{leticiapalhares@gmail.com}\,\,,
\textbf{S.~P.~Sorella}\thanks{silvio.sorella@gmail.com}\,\,\,,\\[2mm]
{\small \textnormal{  \it Departamento de F\'{\i }sica Te\'{o}rica, Instituto de F\'{\i }sica, UERJ - Universidade do Estado do Rio de Janeiro,}}
 \\ \small \textnormal{ \it Rua S\~{a}o Francisco Xavier 524, 20550-013 Maracan\~{a}, Rio de Janeiro, Brasil}\normalsize}
\maketitle

\begin{abstract}
In light of the development of the Gribov issue for pure Euclidean gauge theories and of the recent lattice measurement of soft breaking of the BRST invariance in Yang-Mills theories in the Landau gauge,
we consider non-perturbative features in the gauge-interacting matter sector and their relation with general properties of the Faddeev-Popov operator.
A signature for BRST breaking in the matter sector is proposed and a local and renormalizable framework is constructed, accommodating this signature
and predicting non-perturbative matter propagators that are consistent with available lattice data for adjoint scalars and quarks.
\end{abstract}

\section{Introduction}
Nowadays, the issue of the Gribov copies  \cite{Gribov:1977wm} is an important tool in order to investigate the behavior of nonabelian gauge theories in the non-perturbative infrared region, exhibiting a deep connection with gluon confinement\footnote{See refs.\cite{Sobreiro:2005ec,Vandersickel:2012tz}  for a pedagogical introduction to the Gribov problem. }.  As is widely known, the existence of the Gribov copies is a general feature of the gauge fixing procedure \cite{Singer:1978dk}, reflecting the impossibility of selecting a unique gauge field configuration for each gauge orbit through a local, covariant and renormalizable gauge condition.  \\\\Although a full resolution of the Gribov problem is still lacking, the interplay between analytic methods and numerical lattice  simulations which has taken place during the last decade  has provided strong evidence   for the relevance of the issue of  Gribov copies   in the non-perturbative study of the correlation functions of Euclidean Yang-Mills theories.  A nice example of this fruitful interplay between analytic and numerical methods is provided by the Landau gauge.   If, on one side, several properties of the  Gribov region $\Omega$ of the Landau gauge have been rigorously established from a mathematical point of view   \cite{Dell'Antonio:1989jn,Dell'Antonio:1991xt,vanBaal:1991zw}, on the other side, this gauge possesses a lattice formulation  \cite{Cucchieri:2007md,Cucchieri:2007rg,Cucchieri:2008fc,Cucchieri:2011ig,Cucchieri:2012cb,Oliveira:2012eh,Oliveira:2008uf,Bornyakov:2009ug,Ilgenfritz:2010gu}, which has allowed  for a direct comparison between analytic and numerical results. \\\\  
These great advances in pure-gauge theories have not provided up to now an equivalent development in the understanding of the non-perturbative behavior of gauge-interacting matter. The aim of this paper is to make a series of observations concerning non-perturbative infrared properties of confining theories that also extend to the matter sector. We shall show that a consistent description of confined matter propagators may be achieved through a systematic soft BRST breaking construction, in analogy with what was found for the gauge fields.\\\\ In order to be more precise on the statement of our goals and   for the benefit of the reader, let us give here a short update of the Gribov issue in the Landau gauge. Let us start with the definition of the Gribov region $\Omega$, which is at the basis of the  Gribov-Zwanziger framework \cite{Gribov:1977wm,Zwanziger:1988jt,Zwanziger:1989mf,Zwanziger:1992qr}. The Gribov region $\Omega$ is defined as the set of all gauge  
field configurations fulfilling the Landau gauge condition, $\partial_\mu A^{a}_\mu=0$,  and for which the Faddeev-Popov operator, ${\cal M}^{ab}=-(\partial^2 \delta^{ab} -g f^{abc}A^{c}_{\mu}\partial_{\mu})$, is strictly positive, namely 
\begin{align}
\Omega \;= \; \{ A^a_{\mu}\;; \;\; \partial_\mu A^a_{\mu}=0\;; \;\; {\cal M}^{ab}=-(\partial^2 \delta^{ab} -g f^{abc}A^{c}_{\mu}\partial_{\mu})\; >0 \; \} \;. \label{gr}
\end{align} 
The region $\Omega$ enjoys the following properties  \cite{Dell'Antonio:1989jn,Dell'Antonio:1991xt}:
\begin{enumerate}[i)] 
\item $\Omega$  is convex and bounded in all direction in field space. Its boundary, $\partial \Omega$, is the Gribov horizon, where the first vanishing eigenvalue of the Faddeev-Popov operator shows up.
 \item every gauge orbit crosses at least once the region $\Omega$. 
\end{enumerate} 
In particular, the result { ii)} provides a well defined support to original Gribov's proposal \cite{Gribov:1977wm} of restricting the domain of integration in the functional integral to the region $\Omega$. Therefore, for the partition function of Yang-Mills theories one writes 
\begin{equation}
 Z = \;  \int_\Omega  {\cal D}A\; \delta(\partial A) \; \left( det{\cal M} \right) \;  \; e^{-S_{YM} } = 
\int_\Omega {\cal D}A\;{\cal D}c\;{\cal D}\bar{c}\; {\cal D} b\; e^{-S_{FP}}  \;, \label{gb1}
\end{equation}
where $S_{FP}$ is the Faddeev-Popov action in the Landau gauge
\begin{equation}
S_{FP} = S_{YM} + S_{gf}  \;, \label{fp}
\end{equation}
where $S_{YM}$  and $S_{gf}$ denote, respectively, the Yang-Mills and the gauge-fixing term:
\begin{equation} 
S_{YM} = \frac{1}{4} \int d^{4}x \; F^{a}_{\mu \nu}F^{a}_{\mu\nu} \;, \label{YM}
\end{equation}
and  
\begin{equation}
S_{gf} = \int d^{4}x \left(  b^{a}\partial_{\mu}A^{a}_{\mu}
+\bar{c}^{a} \partial_{\mu}D^{ab}_{\mu}c^{b}  \right) \;,  \label{gf}
\end{equation}
where $({\bar c}^a, c^a)$ are the Faddeev-Popov ghosts, $b^a$ is the Lagrange multiplier implementing the Landau gauge, $D^{ab}_\mu =( \delta^{ab}\partial_\mu + g f^{acb}A^{c}_{\mu})$ is the covariant derivative in the adjoint representation of the gauge group $SU(N)$, and $F^{a}_{\mu\nu}$ denotes the field strength 
\begin{equation}
F^{a}_{\mu\nu} = \partial_{\mu}A^{a}_{\nu} - \partial_{\nu}A^{a}_{\mu} + gf^{abc}A^{b}_{\mu}A^{c}_{\nu}\;. \label{fstr}
\end{equation}
Following \cite{Gribov:1977wm,Zwanziger:1988jt,Zwanziger:1989mf,Zwanziger:1992qr}, the restriction of the domain of integration in the path integral is achieved by adding to the Faddeev-Popov action $S_{FP} $ an additional term $H(A)$, called the horizon term, given by the following non-local expression 
\begin{align}
H(A)  =  {g^{2}}\int d^{4}x\;d^{4}y\; f^{abc}A_{\mu}^{b}(x)\left[ {\cal M}^{-1}\right]^{ad} (x,y)f^{dec}A_{\mu}^{e}(y)   \;,  \label{hf1}
\end{align}
where ${\cal M}^{-1}$ stands for the inverse of the Faddeev-Popov operator. For the partition function one gets \cite{Gribov:1977wm,Zwanziger:1988jt,Zwanziger:1989mf,Zwanziger:1992qr}
\begin{equation}
 Z = \;
\int_\Omega {\cal D}A\;{\cal D}c\;{\cal D}\bar{c}\; {\cal D} b\; e^{-S_{FP}}  =   \int {\cal D}A\;{\cal D}c\;{\cal D}\bar{c} \; {\cal D} b \; e^{-(S_{FP}+\gamma^4 H(A) -V\gamma^4 4(N^2-1))} 
\;, \label{zww1}
\end{equation}
where $V$ is the Euclidean space-time volume. The parameter $\gamma$ has the dimension of a mass and is known as the Gribov parameter. It is not a free parameter of the theory. It is a dynamical quantity, being determined in a self-consistent way through a gap equation called the horizon condition \cite{Gribov:1977wm,Zwanziger:1988jt,Zwanziger:1989mf,Zwanziger:1992qr}, given by 
\begin{equation}
\left\langle H(A)   \right\rangle = 4V \left(  N^{2}-1\right) \;,   \label{hc1}
\end{equation}
where the vacuum expectation value $\left\langle H(A)  \right\rangle$  has to be evaluated with the measure defined by eq.\eqref{zww1}. \\\\Although the horizon term  $H(A)$, eq.\eqref{hf1}, is non-local, it can be cast in local form by means of the introduction of a set of auxiliary fields $(\bar{\omega}_\mu^{ab}, \omega_\mu^{ab}, \bar{\varphi}_\mu^{ab},\varphi_\mu^{ab})$, where $(\bar{\varphi}_\mu^{ab},\varphi_\mu^{ab})$ are a pair of bosonic fields, while $(\bar{\omega}_\mu^{ab}, \omega_\mu^{ab})$ are anti-commuting. It turns out that the partition function $Z_{GZ}$  in eq.\eqref{zww1} can be rewritten as \cite{Zwanziger:1988jt,Zwanziger:1989mf,Zwanziger:1992qr}
\begin{equation}
 Z = \;
\int {\cal D}A\;{\cal D}c\;{\cal D}\bar{c}\; {\cal D} b \; {\cal D}{\bar \omega}\; {\cal D} \omega\; {\cal D} {\bar \varphi} \;{\cal D} \varphi \; e^{-S_{GZ}} \;, \label{lzww1}
\end{equation}
where $S_{GZ}$ is given by the local expression 
\begin{equation} 
S_{GZ} = S_{YM} + S_{gf} + S_0+S_\gamma  \;, \label{sgz}
\end{equation}
with
\begin{equation}
S_0 =\int d^{4}x \left( {\bar \varphi}^{ac}_{\mu} (\partial_\nu D^{ab}_{\nu} ) \varphi^{bc}_{\mu} - {\bar \omega}^{ac}_{\mu}  (\partial_\nu D^{ab}_{\nu} ) \omega^{bc}_{\mu}  - gf^{amb} (\partial_\nu  {\bar \omega}^{ac}_{\mu} ) (D^{mp}_{\nu}c^p) \varphi^{bc}_{\mu}  \right) \;, \label{s0}
\end{equation}
and 
\begin{equation}
S_\gamma =\; \gamma^{2} \int d^{4}x \left( gf^{abc}A^{a}_{\mu}(\varphi^{bc}_{\mu} + {\bar \varphi}^{bc}_{\mu})\right)-4 \gamma^4V (N^2-1)\;. \label{hfl}
\end{equation} 
In the local formulation of the Gribov-Zwanziger action, the horizon condition \eqref{hc1} takes the simpler form 
\begin{equation}
 \frac{\partial \mathcal{E}_v}{\partial\gamma^2}=0\;,   \label{ggap}
\end{equation}
where $\mathcal{E}_{v}(\gamma)$ is the vacuum energy defined by:
\begin{equation}
 e^{-V\mathcal{E}_{v}}=\;Z\;  \label{vce} \;.
\end{equation}
The local action $S_{GZ}$ in eq.\eqref{sgz} is known as the Gribov-Zwanziger action. It has been shown to be renormalizable to all orders \cite{Zwanziger:1988jt,Zwanziger:1989mf,Zwanziger:1992qr}. \\\\Recently, a refinement of the Gribov-Zwanziger action has been worked out by the authors  \cite{Dudal:2007cw,Dudal:2008sp,Dudal:2011gd}, by taking into account the existence of certain dimension two condensates\footnote{See \cite{Gracey:2010cg,Thelan:2014mza} for a recent detailed investigation on  the structure of these condensates in color space.}.  The Refined Gribov-Zwanziger (RGZ) action reads \cite{Dudal:2007cw,Dudal:2008sp,Dudal:2011gd}
\begin{equation}
S_{RGZ} = S_{GZ} + \int d^4x \left(  \frac{m^2}{2} A^a_\mu A^a_\mu  - \mu^2 \left( {\bar \varphi}^{ab}_{\mu}  { \varphi}^{ab}_{\mu} -  {\bar \omega}^{ab}_{\mu}  { \omega}^{ab}_{\mu} \right)   \right)  \;,  \label{rgz}
\end{equation}
where $S_{GZ}$ stands for the Gribov-Zwanziger action,  eq.\eqref{sgz}.  As much as the Gribov parameter $\gamma^2$, the massive parameters $(m^2, \mu^2)$ have a dynamical origin, being related to the existence of the dimension two condensates $\langle A^a_\mu A^a_\mu \rangle$ and  $\langle {\bar \varphi}^{ab}_{\mu}  { \varphi}^{ab}_{\mu} -  {\bar \omega}^{ab}_{\mu}  { \omega}^{ab}_{\mu}  \rangle$, \cite{Dudal:2007cw,Dudal:2008sp,Dudal:2011gd}. 
The gluon propagator obtained from the RGZ action turns out to be suppressed in the infrared region, attaining a non-vanishing value at zero momentum, $k^2=0$, {\it i.e.}
\begin{eqnarray} 
\langle  A^a_\mu(k)  A^b_\nu(-k) \rangle  & = &  \delta^{ab}  \left(\delta_{\mu\nu} - \frac{k_\mu k_\nu}{k^2}     \right)   {\cal D}(k^2) \;, \label{glrgz} \\
{\cal D}(k^2) & = & \frac{k^2 +\mu^2}{k^4 + (\mu^2+m^2)k^2 + 2Ng^2\gamma^4 + \mu^2 m^2}  \;. \label{Dg}
\end{eqnarray} 
Also, unlike the case of the GZ action, the ghost propagator stemming from the Refined theory is not enhanced in the deep infrared:
\begin{equation}
{\cal G}^{ab}(k^2) = \langle  {\bar c}^{a} (k)  c^b(-k) \rangle \Big|_{k\sim 0} \; \sim \frac{\delta^{ab}}{k^2}   \;.\label{ghrgz} 
\end{equation}
The infrared behaviour of the  gluon and ghost propagators obtained from the RGZ  action turns out to be in very good agreement with the most recent  numerical lattice simulations on large lattices \cite{Cucchieri:2007rg,Cucchieri:2008fc,Cucchieri:2011ig}. Moreover, the numerical estimates  \cite{Cucchieri:2011ig}  of the parameters $(m^2,\mu^2,\gamma^2)$ show that the RGZ gluon propagator \eqref{glrgz} exhibits complex poles and violates  reflection positivity. This kind of two-point function lacks the  K{\"a}ll{\'e}n-Lehmann spectral representation and cannot be associated with the propagation of physical particles. Rather, it indicates that, in the non-perturbative infrared region, gluons are not physical excitations of the spectrum of the theory, {\it i.e.} they are confined.  It is worth mentioning here that the RGZ gluon propagator has been employed in analytic calculation of the first glueball states  \cite{Dudal:2010cd,Dudal:2013wja}, yielding results which compare well with the available numerical simulations as well as with other approaches, see \cite{Mathieu:2008me} for an  account on this topic. The RGZ gluon propagator has also been used in order to  study the Casimir energy within the MIT bag model \cite{Canfora:2013zna}. The resulting energy has the correct expected confining behaviour. Applications  of the RGZ theory at finite temperature can be found in  \cite{Fukushima:2013xsa,Canfora:2013kma}. In \cite{Capri:2012ah,Capri:2013oja}, the issue of the Gribov copies has been addressed in the case in which Higgs fields are present, yielding analytic results on  the hard problem of the understanding of the transition between the confining and Higgs phases for asymptotically free gauge theories. The output of this analysis  turns out to be in qualitative agreement with the seminal work by  Fradkin-Shenker \cite{Fradkin:1978dv}. Finally, in \cite{Capri:2014xea,Capri:2014tta}, the Gribov-Zwanziger construction has been generalised to supersymmetric Yang-Mills theories. All these results enable us to state that the issue of the Gribov copies  captures nontrivial aspects of the non-perturbative dynamics of Yang-Mills theories. \\\\One important aspect of both GZ and RGZ theories is that they exhibit a soft breaking of the BRST symmetry. In fact, introducing the nilpotent BRST transformations 
\begin{eqnarray}
\label{brst1}
sA^{a}_{\mu} &=& - D^{ab}_{\mu}c^{b}\;,\nonumber \\
s c^{a} &=& \frac{1}{2}gf^{abc}c^{b}c^{c} \;, \nonumber \\
s{\bar c}^{a} &=& b^{a}\;, \qquad \; \; 
sb^{a} = 0 \;, \nonumber \\
s{\bar \omega}^{ab}_\mu & = & {\bar \varphi}^{ab}_\mu \;, \qquad  s {\bar \varphi}^{ab}_\mu =0\;, \nonumber \\
s { \varphi}^{ab}_\mu&  = & {\omega}^{ab}_\mu  \;, \qquad s {\omega}^{ab}_\mu = 0 \;, 
\end{eqnarray}
it is immediately checked that the Gribov-Zwanziger action breaks the BRST symmetry, as summarized by the equation\footnote{A similar equation holds in the case of the RGZ action \cite{Dudal:2007cw,Dudal:2008sp,Dudal:2011gd}.}
\begin{equation}
s S_{GZ} = \gamma^2 \Delta  \;, \label{brstbrr}
\end{equation}
where 
\begin{equation}
\Delta = \int d^{4}x \left( - gf^{abc} (D_\mu^{am}c^m) (\varphi^{bc}_{\mu} + {\bar \varphi}^{bc}_{\mu})   + g f^{abc}A^a_\mu \omega^{bc}_\mu            \right)  \;. \label{brstb1}
\end{equation}
Notice that the breaking term $\Delta$ is of dimension two in the fields. As such, it is a soft breaking. The properties of the soft breaking of the BRST symmetry of the Gribov-Zwanziger theory and its relation with confinement have been object of intensive investigation, see  \cite{Baulieu:2008fy,Dudal:2009xh,Sorella:2009vt,Sorella:2010it,Capri:2010hb,Dudal:2012sb,Reshetnyak:2013bga}. Let us mention here that the broken identity  \eqref{brstbrr} is deeply connected with the restriction to the Gribov region $\Omega$. Equation \eqref{brstbrr} can be translated into a set of softly broken Slavnov-Taylor  identities which ensure the all order renormalizability of both GZ and RGZ actions. The presence of the soft breaking term $\Delta$ turns out to be necessary in order to have a confining gluon propagator which attains a non-vanishing value at zero momentum, eqs.\eqref{glrgz},\eqref{Dg}, in agreement with the lattice data \cite{Cucchieri:2007rg,Cucchieri:2008fc,Cucchieri:2011ig}. It is worth underlining that this  property is deeply related to the soft breaking of the BRST symmetry. In fact, the non-vanishing of the propagator at zero momentum relies on  the parameter $\mu^2$, which reflects the existence of the   BRST-exact dimension-two condensate \cite{Dudal:2007cw,Dudal:2008sp,Dudal:2011gd}
\begin{equation}
\langle {\bar \varphi}^{ab}_{\mu}  { \varphi}^{ab}_{\mu} -  {\bar \omega}^{ab}_{\mu}  { \omega}^{ab}_{\mu}  \rangle  = \langle s ( {\bar \omega}^{ab}_{\mu}(x) { \varphi}^{ab}_{\mu}(x) )  \rangle \neq 0 \;. \label{conda} 
\end{equation}
Moreover, despite the soft breaking, eq.\eqref{brstbrr}, a set of BRST  invariant composite operators whose correlation functions exhibit the K{\"a}ll{\'e}n-Lehmann spectral representation with positive spectral densities can be consistently introduced \cite{Baulieu:2009ha}. \\\\Although a satisfactory understanding of the physical meaning of the soft breaking of the BRST symmetry in presence of the Gribov horizon and of its relationship with confinement is still lacking, it is worth  underlining here that the first concrete numerical lattice evidence  of the existence of such breaking has been provided by the authors   of \cite{Cucchieri:2014via},  who have shown that a BRST exact correlation function is non-vanishing, signaling thus the breaking of the BRST symmetry. More precisely, in \cite{Cucchieri:2014via}, the infrared behaviour of the correlation function 
\begin{eqnarray} 
{\cal Q}^{abcd}_{\mu\nu}(x-y)  & = &  \langle {\cal R}^{ab}_\mu(x)  {\cal R}^{cd}_\nu(y)  \rangle    \;, \label{rr} \\[3mm]
{\cal R}^{ac}_\mu(x) & = &  \int d^4z\;  ({\cal M}^{-1})^{ad} (x,z) \; g f^{dec} A^{e}_\mu(z)  \;, \label{ra} 
\end{eqnarray} 
  involving the inverse of the Faddeev-Popov operator $\cal M$,
has been investigated through numerical lattice simulations. The relation of the correlation function \eqref{rr} with the breaking of the BRST symmetry can be understood by observing that, within the local formulation of the Gribov-Zwanziger framework, expression \eqref{rr}  corresponds to the exact correlation function 
\begin{equation} 
\langle \; s ( \varphi^{ab}_\mu(x) {\bar \omega}^{cd}_\nu(y)  \; ) \rangle  = \langle   \omega^{ab}_\mu(x) {\bar \omega}^{cd}_\nu(y) + \varphi^{ab}_\mu(x) {\bar \varphi}^{cd}_\nu(y)      \rangle \;.  \label{ss}
\end{equation}
In fact, integrating out the auxiliary fields $(\bar{\omega}_\mu^{ab}, \omega_\mu^{ab}, \bar{\varphi}_\mu^{ab},\varphi_\mu^{ab})$ in expression 
\begin{equation}
\int [{\cal D} {\Phi}] \; \left( \omega^{ab}_\mu(x) {\bar \omega}^{cd}_\nu(y) + \varphi^{ab}_\mu(x) {\bar \varphi}^{cd}_\nu(y) \right)   \; e^{-S_{GZ}} \;,  \label{loce}
\end{equation}
gives
\begin{equation} 
\frac{ \int [{\cal D} {\Phi}] \;   \left( s \left( \varphi^{ab}_\mu(x) {\bar \omega}^{cd}_\nu(y)  \right)   \right) \; e^{-S_{GZ}} }{ \int [{\cal D \phi}]    \; e^{-S_{GZ}}}  =\gamma^4 \; \frac{  \int {\cal D}A\; \delta(\partial A) \; \left( det{\cal M} \right) \; {\cal R}^{ab}_\mu(x)  {\cal R}^{cd}_\nu(y)  \; e^{-(S_{YM}+\gamma^4 H(A) )} } {\int {\cal D}A\; \delta(\partial A) \; \left( det{\cal M} \right)   \; e^{-(S_{YM}+\gamma^4 H(A) )}  } \;. \label{brstbr}
\end{equation}
This equation shows  that the investigation of the correlation function \eqref{rr} with a cutoff at the Gribov horizon is directly related to the existence of the BRST breaking. This is precisely what has been done in \cite{Cucchieri:2014via}, where the correlator \eqref{rr} has been shown to be non-vanishing, see Fig.1 of \cite{Cucchieri:2014via}. Moreover, from \cite{Cucchieri:2014via}, it turns out that in the deep infrared the Fourier transform of the correlation function \eqref{rr} is deeply enhanced, see Fig.2 of \cite{Cucchieri:2014via}, behaving as $\frac{1}{k^4}$, namely 
\begin{equation} 
 \langle {\tilde {\cal R}} ^{ab}_\mu(k)  {\tilde {\cal R}}^{cd}_\nu(-k)  \rangle  \sim \frac{1}{k^4} \;.  \label{enhanc}
\end{equation}  
As observed in \cite{Cucchieri:2014via}, this behaviour can be  understood by making use of the analysis \cite{Zwanziger:2010iz}, {\it i.e.} of the cluster decomposition 
\begin{equation}
 \langle {\tilde {\cal R}} ^{ab}_\mu(k)  {\tilde {\cal R}}^{cd}_\nu(-k)  \rangle   \sim  g^2 {\cal G}^2(k^2) {\cal D}(k^2) \;, \label{clust} 
\end{equation} 
where ${\cal D}(k^2)$ and ${\cal G}(k^2)$ correspond to the   gluon and ghost propagators, eqs.\eqref{Dg},\eqref{ghrgz}. A non-enhanced ghost propagator, {\it i.e.}  ${\cal G}(k^2) \Big|_{k\sim 0} \sim \frac{1}{k^2}$, and an infrared finite gluon propagator, {\it i.e.} ${\cal D}(0) \neq 0$, nicely yield the behaviour of eq.\eqref{enhanc}. \\\\The aim of the present work is that of showing that the quantity ${\cal R}$, eq.\eqref{ra}, and the correlation function  $ \langle {\cal R}(x)  {\cal R}(y)  \rangle $, eq.\eqref{rr}, can be consistently generalised to the case of matter fields, {\it i.e.} when quark and scalar fields are included in the starting action. \\\\More precisely, let $F^{i}$ denote a generic matter field in a  given representation of $SU(N)$, specified by the generators $(T^a)^{ij}$, $a=1,..,(N^2-1)$, and let  ${\cal R}^{ai}(x)$ stand for the quantity
\begin{equation}
{\cal R}^{ai}(x)  =  g \int d^4z\;  ({\cal M}^{-1})^{ab} (x,z)   \;(T^b)^{ij} \;F^{j}(z)   \label{rmatter}  \;, 
\end{equation}
which is a convolution of the inverse Faddeev-Popov operator with a given colored matter field, being clearly the matter counterpart of the operator ${\cal R}^{ab}_{\mu}$ in the pure gauge case.
We shall be able to prove that, in analogy with the case of the gauge field $A^a_\mu$, a   non-trivial correlation function 
\begin{equation} 
 \langle {\cal R}^{ai}(x)  {\cal R}^{bj}(y)  \rangle    \;, \label{rcm} 
\end{equation} 
can be obtained from a local and renormalizable action  which is constructed by adding to the starting conventional matter action a non-local term which shares great similarity with the horizon function $H(A)$, eq.\eqref{hf1}, namely 
\begin{equation}
{g^{2}}   \int d^{4}x\;d^{4}y\; F^{i}(x) (T^a)^{ij} \left[ {\cal M}^{-1}\right]^{ab} (x,y) (T^b)^{jk} F^{k} (y) \;.     \label{hmatter}
\end{equation} 
As it happens in the case of the Gribov-Zwanziger theory, the term \eqref{hmatter} can be cast in local form by means of the introduction of suitable auxiliary fields. The resulting local action enjoys a large set of Ward identities which guarantee its renormalizabilty. The introduction of the term \eqref{hmatter} deeply modifies the infrared behaviour of the correlation functions of the matter fields giving rise, in particular, to propagators which are of the confining type, while being in good agreement with the available lattice data, as in the case of the scalar matter fields  \cite{Maas:2011yx,Maas:2010nc}   as well as in the case of quarks   \cite{Furui:2006ks,Parappilly:2005ei}. \\\\Moreover, relying on the  numerical data for the two-point correlation functions of quark and scalar fields, expression \eqref{rcm} turns out to be non-vanishing and, interestingly enough, it seems to behave exactly as expression \eqref{enhanc} in the  deep infrared, {\it i.e.} 
\begin{equation} 
 \langle {\tilde {\cal R}} ^{ai}(k)  {\tilde {\cal R}}^{bj}(-k)  \rangle  \sim \frac{1}{k^4} \;.  \label{menhanc}
\end{equation} 
Also, as in the case of the gauge sector, expression \eqref{rcm} signals the existence of the BRST breaking in the matter field sector of the theory. \\\\The present work is organized as follows. In Sect.2 we present a discussion of the correlation function \eqref{rcm}  in the case of  quark and scalar fields, relying on the available data for the quark and scalar propagators. In Sect.3 we shall show how the correlation function $\langle {\cal R}^{ai}(x)  {\cal R}^{bj}(y)  \rangle $ can be obtained from a local and renormalizable action exhibiting a soft breaking of the BRST invariance in the matter sector. This will be done by working out in detail the case of a scalar field in the adjoint representation, in Subsect.3.1. We shall also discuss how $\langle {\cal R}^{ai}(x)  {\cal R}^{bj}(y)  \rangle $ encodes information on the soft  breaking of the BRST symmetry. In Subsect.3.2 we generalize the previous construction  to the case of quark fields. Sect.4  contains  our conclusion. The final Appendix collects  the details of the algebraic proof of the renormalizability of the local action obtained by the addition of the term  \eqref{hmatter} in the case of a scalar matter field in the adjoint representation.  

\section{Discussion on the correlation function $\langle {\tilde {\cal R}}(k) {\tilde {\cal R}}(-k) \rangle$ from the available lattice data on the propagators of  scalar and quark fields}

Let us now investigate the correlation function $\langle {\tilde {\cal R}}(k) {\tilde {\cal R}}(-k) \rangle$, that signals soft BRST breaking in the matter sector, in  light of available lattice data for gauge-interacting matter propagators in the Landau gauge.

As in the pure gauge case, one may rely on the general cluster decomposition property in order to obtain the leading behavior in the deep infrared region. With this aim, one writes the ${\cal R}^{ai}(x)$ function in terms of elementary fields, with the inverse Faddeev-Popov propagator represented via ghost fields as usual:
\begin{eqnarray}
\langle {\cal R}^{ai}(x) {\cal R}^{bj}(y)
\rangle
&=&
{g^{2}}   \int d^{4}z\;d^{4}z'\; \langle \bar{c}^{a}(x) c^{a'}(z) F^{i'}(z) (T^{a'})^{i'i}  \bar{c}^b(x) c^{b'}(z') (T^{b'})^{j'j} F^{j'} (z') 
\rangle
\\
&=& g^2  (T^{a})^{i'i} (T^{b})^{i'j}\int d^4k {\rm e}^{ik(x-y)}{\cal G}^2(k)D(k^2)+
\nonumber\\
&&+
{g^{2}}   \int d^{4}z\;d^{4}z'\; \langle \bar{c}^{a}(x) c^{a'}(z) F^{i'}(z) (T^{a'})^{i'i}  \bar{c}^b(x) c^{b'}(z') (T^{b'})^{j'j} F^{j'} (z') 
\rangle_{1PI}
\;,
\end{eqnarray}
where ${\cal G}(k^2)$ is the ghost propagator, while $D(k^2)$ now stands for the propagator of the associated matter field. The one-particle-irreducible (1PI) contribution above becomes subleading in the IR limit, since in this case the points $x$ and $y$ are largely separated and the cluster decomposition applies. This can also be seen diagrammatically.
Since the external legs are ghosts, these corrections will involve at least two ghost-gluon vertices, that carry a derivative coupling. In fact, as a consequence of the transversality 
of the gluon propagator, factorization of the external momentum takes place, implying the subleading character of the 1PI contributions.

Therefore, in the limit $k\to 0$, the (full) ghost and matter propagators alone dictate the momentum-dependence of the correlation function $\langle {\tilde {\cal R}}(k) {\tilde {\cal R}}(-k) \rangle$, i.e.
\begin{eqnarray}
\langle \tilde{\cal R}^{ai}(k) \tilde{\cal R}^{bj}(-k)
\rangle
&\sim& g^2 {\cal G}^2(k)D(k^2)
\;.
\end{eqnarray}
Having in mind the non-enhanced ghost propagator, ${\cal G}(k^2)\sim 1/k^2$ (as observed in high-precision pure gauge simulations in the Landau gauge \cite{Cucchieri:2007rg,Cucchieri:2008fc,Cucchieri:2011ig}), it is straightforward to conclude that a finite zero-momentum value for the matter propagators is a sufficient condition for a $\sim 1/k^4$  behavior of the correlation function $\langle {\tilde {\cal R}}(k) {\tilde {\cal R}}(-k) \rangle$ in the deep IR.

As we shall see in the following subsections, both scalar and fermion propagators display, when coupled to non-Abelian gauge fields, a shape compatible with a finite zero-momentum value in the currently available lattice data. We expect thus a $\sim 1/k^4$ behavior of  the correlation function $\langle {\tilde {\cal R}}(k) {\tilde {\cal R}}(-k) \rangle$ in the matter sector, being in this sense a universal property associated with the Faddeev-Popov operator -- when coupled to any colored field --  in confining Yang-Mills theories that can be easily probed in the future via direct lattice measurements.

Moreover, fits of the lattice data are presented for adjoint scalars in Subsect.\ref{scalars} and for fermions in Subsect.\ref{quarks}. This analysis shows that the propagators for gauge-interacting scalars and fermions are compatible not only with a finite zero-momentum limit, but also with a complete analytical form that can be extracted from an implementation of soft BRST breaking in the matter sector to be presented below, in Sect.3. 

\subsection{ The case of the scalar field in the adjoint representation}\label{scalars}

In this subsection, we consider real scalar fields coupled to a confining Yang-Mills theory:
\begin{eqnarray}
{\cal L} = \frac{1}{4} F_{\mu\nu}^aF_{\mu\nu}^a +\frac{1}{2} [D_{\mu}^{ab}\phi^b]^2 + \frac{m_{\phi}^2}{2}\phi^a\phi^a +\frac{\lambda}{4!} [\phi^a\phi^a]^2  + {\cal L}_{GF}\;,
\end{eqnarray}
where  ${\cal L}_{GF}$ is the Landau gauge fixing term and $\phi$ is a real scalar field in the adjoint representation of $SU(N)$ and there is no Higgs mechanism, namely $\langle\phi\rangle = 0$.  

We are interested in analyzing the infrared non-perturbative regime, focussing especially on the adjoint scalar propagator. We resort to the lattice implementation of this system: currently available in the quenched approximation with the specific setup described in \cite{Maas:2010nc}. Preliminary and unpublished data points for larger lattice sizes (with lattice cutoff $a^{-1}=4.94 $ GeV and $N=30$ lattice sites) \cite{axel} are displayed in Fig. 1 for different values of the bare scalar mass ($m_{bare}=0,\,1,\,10$ GeV). It should be noticed that this data is unrenormalized in the lattice sense. The renormalization procedure that fixes the data to a known renormalization scheme and the resulting points will be discussed below.

\begin{figure}[h!]
\center
\includegraphics[width=9cm]{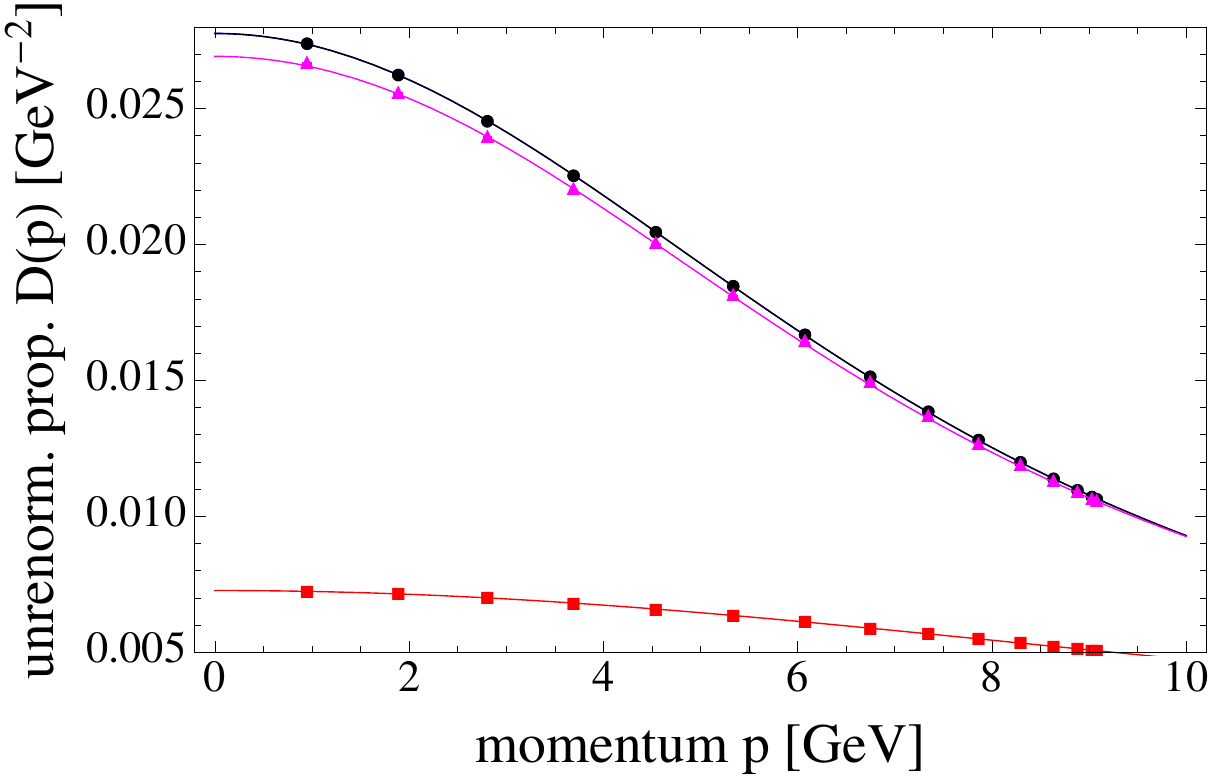}
\caption{Unrenormalized propagator for different bare masses of the scalar field: $m_{bare}=0$(top, black), 1 and 10 GeV (bottom, red). The points are preliminary and unpublished lattice data from quenched simulations \cite{axel} (for lattice cutoff $a^{-1}=4.54 $ GeV, $N=30$ and $\beta=2.698$; cf. also \cite{Maas:2010nc} for more details on the lattice setup and measurements) and the curves are the corresponding fits, whose parameter values can be found in Table 1.}
\end{figure}

First of all, the data tends to show a finite zero-momentum value for the scalar propagator, irrespective of its bare mass. This indicates   -- together with the well-stablished non-enhanced ghost propagator -- that the correlation function $\langle \tilde{R} \tilde R\rangle_k$ is indeed non-vanishing in the IR limit, presenting the power-law enhancement $\sim1/k^4$ that we have anticipated above.

The curves in Fig. 1 further show that the data is compatible with 4-parameter fits of the following form:
\begin{eqnarray}
D(p)&=& Z\, \frac{p^2+\mu_{\phi}^2}{p^4+p^2(m_{\phi}^2+\mu_{\phi}^2)+\sigma^4+m_{\phi}^2\mu_{\phi}^2}
\,,\label{RGZfit}
\end{eqnarray}
where $Z,\mu_{\phi},m_{\phi},\sigma$ are the fit parameters, whose values are presented in Table 1. In this case we may extrapolate the fits in order to obtain the specific values at zero momentum: $D(p=0)\approx 0.028,\, 0.027,\, 0.0073$ GeV$^{-2}$ for $m=0,1,10$ GeV, respectively, so that the non-trivial IR limit is clear. Moreover, the $\sigma$ parameter -- which will be directly related to the realization of a $\langle{\cal R R}\rangle \ne 0$ in the framework of the next Section -- seems to be nonvanishing.  It is also interesting to point out that the obtained fits correspond to a combination of two complex-conjugate poles for all values of bare scalar mass, indicating the absence of a K\"all\'en-Lehmann spectral representation for this two-point function and the presence of positivity violation. In this sense the adjoint scalar propagators  consistently represent confined degrees of freedom, that do not exhibit a physical propagating pole.
    
    \begin{table}[ht]
 \caption{Fit parameters for the unrenormalized propagator in powers of GeV.}
 \vspace{0.3cm}
  \centering
   \begin{tabular}{c ||c| c| c| c||c}
    $m_{bare}$ & $\mu_{\phi}^2$ & $m_{\phi}^2$& $\sigma^4$ & $Z$&$\chi^2/\textrm{dof}$\\
    \hline\hline 
    0 & 120
    & 0 & 4913 & 1.137 &0.31\\
      \hline 
    1 & 46 & 34  & 644 & 1.28 & 1.84\\
      \hline 
    10 & 88
    & 158 & 1267 & 1.26 & 0.10
     \end{tabular}
     \label{table:par} 
    \end{table}
    
An important issue to be addressed is the possibility of scheme dependence of those findings. To check for this, we have also analyzed the scalar propagators after renormalization in another scheme.
As usual, renormalization is implemented through the inclusion of mass $\delta m_{\phi}$ and wave-function renormalization $\delta Z $ counterterms:
\begin{eqnarray}
D_{ren}^{-1}(p)&=& D^{-1}(p) +\delta m_{\phi}^2 +\delta Z (p^2+m_{bare}^2)
\,,
\end{eqnarray}
where the counterterms are obtained by imposing the following renormalization conditions (for $\Lambda=2$ GeV):
\begin{enumerate}[i)]
\item $\partial_{p^2}D_{ren}^{-1}(p=\Lambda)=1 $;
\item $D_{ren}^{-1}(p=\Lambda)=\Lambda^2+m_{bare}^2$.
\end{enumerate}
The fit functions were used to compute the counterterms and the renormalized points are obtained from the original lattice data by adding the same counterterms\footnote{Direct renormalization of lattice data was avoided, since we did not have access to the measurement of $\partial_{p^2}D$ and the number of data points available was not sufficient for a reliable numerical derivative to be computed.}. Results are shown in Fig. 2 and Table 2.

\begin{figure}[h!]
\center
\includegraphics[width=10cm]{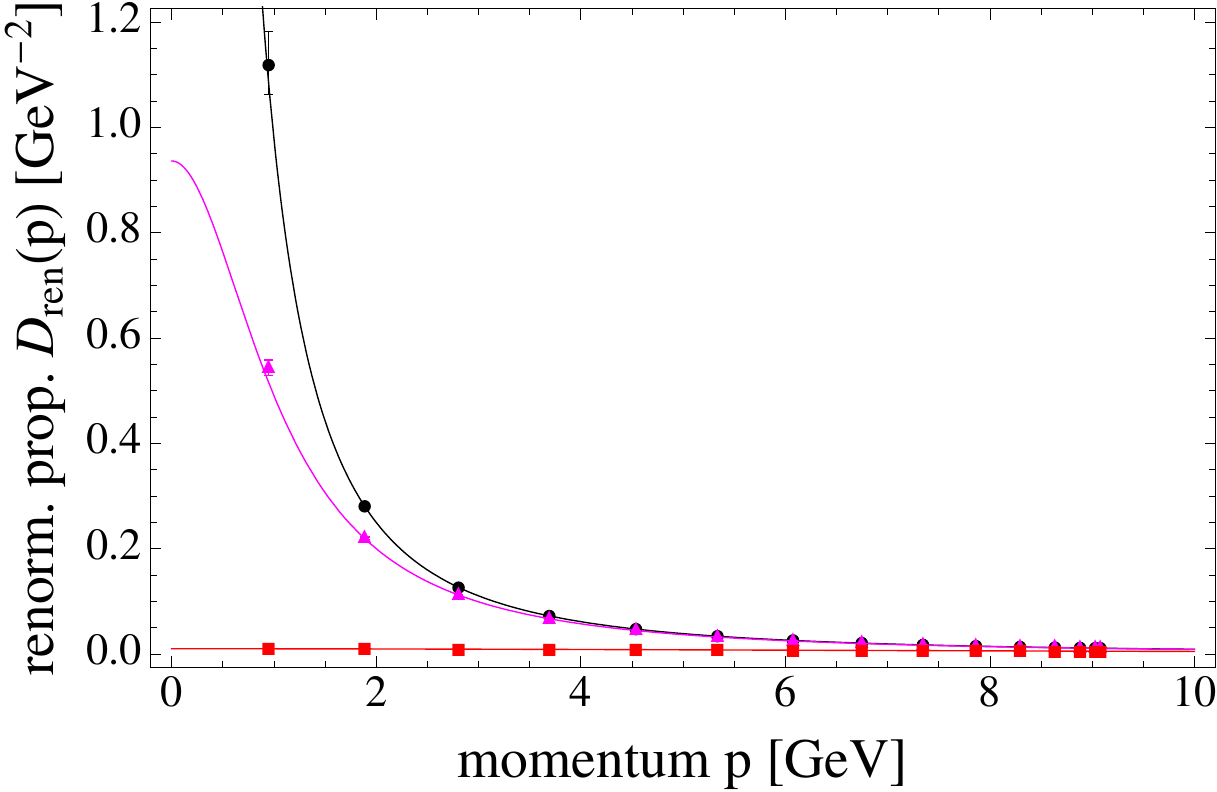}
\caption{Renormalized propagator for different bare masses of the scalar field: $m_{bare}=0$(top, black), 1 and 10 GeV (bottom, red). The points are obtained from the unrenormalized lattice data  \cite{axel,Maas:2010nc} displayed in Fig.1.}
\end{figure}

The renormalized propagator may be rewritten in the form \eqref{RGZfit}, with redefined parameters $m_{\phi}',\sigma',Z'$:
\begin{eqnarray}
D_{ren}(p)&=&
Z'\, \frac{p^2+\mu_{\phi}^2}{p^4+p^2(m_{\phi}'^2+\mu_{\phi}^2)+\sigma'^4+m_{\phi}'^2\mu_{\phi}^2}
\end{eqnarray}    
    
        \begin{table}[ht]
 \caption{Counterterms, redefined fit parameters and zero-momentum values of the renormalized propagator in powers of GeV.}
 \vspace{0.3cm}
  \centering
   \begin{tabular}{c ||c| c|| c| c|c||c}
    $m_{bare}$ & $\delta m_{\phi}^2$ & $\delta Z$& $m_{\phi}'^2$ & $\sigma'^4$ & $Z'$& $D_{ren}(p=0)$\\
   \hline\hline 
    0 & -35.98    & 0.40 & -28.09 & 3374.32 & 0.781&26.7\\
      \hline 
    1 & -36.49 & 0.416 & -8.18  & 420.84 & 0.834 &0.94\\
      \hline 
    10 &  -69.69    & 0.322 & 79.19 & 902.23 & 0.894 &0.01
     \end{tabular}
     \label{table:CT} 
    \end{table}

All the interesting qualitative properties observed in the unrenormalized data remain valid, namely: $(i)$ finite IR limit, $(ii)$ compatibility with 4-parameter fits of the same form, with non-trivial $\sigma$ values, $(iii)$ the fit parameters yield complex-conjugate poles, so that the renormalized propagator is still compatible with positivity violation and confinement.

We underline that the present analysis for the scalar fields is meant to be a preliminary study of the propagator. As such, the results are still at the qualitative level. A more quantitative analysis would require further simulations with improved statistics and even larger lattices.

\subsection{The case of the quark field}\label{quarks}

In this subsection, we consider the case of gauge-interacting fermionic fields coupled to a confining Yang-Mills theory. Of course, the case of QCD is the emblematic example. 
We will verify that the same qualitative properties shown above for scalar fields can also be found in this case, indicating that the IR enhancement of the correlation function $\langle\tilde {\cal R}\tilde {\cal R}\rangle\sim 1/k^4$ seems to be universally present in the confined matter sector.

The fermionic propagator is decomposed as usual,
\begin{eqnarray}
{\cal S}(p)=Z(p^2)\frac{-ip_{\mu}\gamma_{\mu}+{\cal A}(p^2)}{p^2+{\cal A}(p^2)}\,,
\end{eqnarray}
and our interest resides solely on the mass function ${\cal A}(p^2)$, whose lattice data will be analyzed here.

\begin{figure}[h!]
   \centering
       \includegraphics[width=9cm]{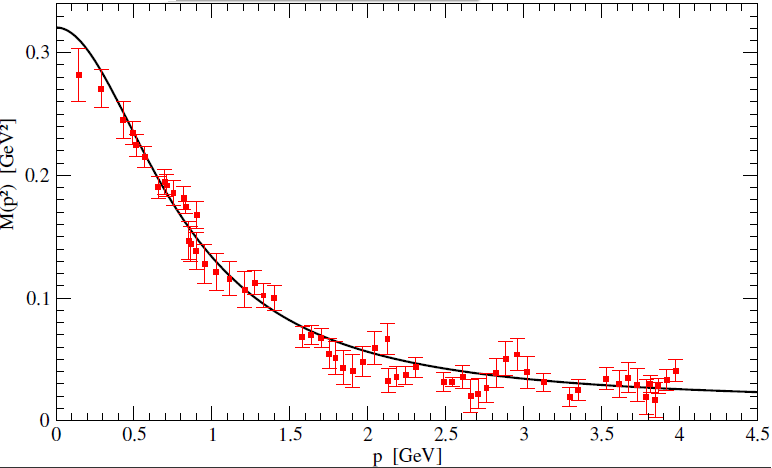}
               \caption{Lattice quark mass function \cite{Parappilly:2005ei} with its fit ${\cal A}(p^2)$. Figure extracted from \cite{Dudal:2013vha}; fit obtained by O. Oliveira \cite{Orlando}.}
\end{figure}

As already discussed and shown in \cite{Dudal:2013vha}, the data of \cite{Parappilly:2005ei} for the mass function of the propagator of degenerate up ($u$) and down ($d$) quarks with current mass $\mu=0.014~\text{GeV}$ can be fitted excellently with
\begin{equation}\label{fit}
{\cal A}(p^2)=\frac{M^3}{p^2+m^2}+\mu~\text{with}~M^3=0.1960(84)~\text{GeV}^3\,, m^2=0.639(46)~\text{GeV}^2 \quad (\chi^2/\text{d.o.f.}~=~1.18)\,.
\end{equation}
as can be seen in Fig. 3. A recent alternative semi-analytic description of the non-perturbative quark propagator in the Landau gauge based on an effective gluon mass was discussed in \cite{Pelaez:2014mxa}.
 
The quark propagator presents clearly a finite IR limit. This is, in fact,  well-known in QCD as dynamical mass generation and is intimately related to chiral symmetry breaking. Interestingly enough, this is also a sufficient condition -- supposing a non-enhanced ghost propagator -- for the soft BRST breaking in the quark sector through the IR enhancement of the correlation function $\langle \tilde{\cal R}\tilde{\cal R}\rangle$. Again, we predict a $\sim 1/k^4$ IR scaling for this observable, now in the quark sector. This suggests a close relation between soft BRST breaking and chiral symmetry breaking, and may provide an interesting underlying connection between confinement and chiral symmetry breaking.

\section{Implementing the correlation function $\langle {\tilde {\cal R}}(k) {\tilde {\cal R}}(-k) \rangle$ within a local quantum field theory framework} 

Now that we have established that available lattice data for different propagators of gauge-interacting matter seems to be qualitatively compatible with the non-trivial $\langle {\tilde {\cal R}}(k) {\tilde {\cal R}}(-k) \rangle\sim 1/k^4$ behaviour, let us discuss how the correlation function \eqref{rcm} can be obtained through a local and renormalizable action. In this section, the example of a real scalar field $\phi^a$ in the adjoint representation of the gauge group will be worked out in detail. 

\subsection{Scalar field in the adjoint representation } 

We start by considering the following non-local action  
\begin{equation}
\label{acs}
S^{\phi} = \int d^4x\; \left(
 \frac{1}{2}(D^{ab}_{\mu}\phi^{b})^{2} + \frac{m^2_{\phi}}{2} \phi^a \phi^a 
+ \frac{\lambda}{4!}(\phi^{a}\phi^{a})^{2} \right)   + {g^{2}} \sigma^4  \int d^{4}x\;d^{4}y\; f^{abc}\phi^{b}(x)\left[ {\cal M}^{-1}\right]^{ad} (x,y)f^{dec}\phi^{e}(y) \;, 
\end{equation}
where $\sigma$ is a massive parameter which, to some extent, plays a role akin to that of the Gribov parameter $\gamma^2$ of the Gribov-Zwanziger action. eq.\eqref{sgz}.\\\\Following now the same procedure adopted in the case of the Gribov-Zwanziger action, it is not difficult to show that the non-local action \eqref{acs} can be cast in local form. This is achieved by introducing  a set of auxiliary fields $(\tilde{\eta}^{ab},\eta^{ab})$, $(\tilde{\theta}^{ab},\theta^{ab})$, where $(\tilde{\eta}^{ab},\eta^{ab})$ are commuting fields while  $(\tilde{\theta}^{ab},\theta^{ab})$ are anti-commuting. For the local version of  \eqref{acs} one gets 
\begin{equation}
S_{loc}^{\phi}   =     S_{0}^{\phi} + S_{\sigma}   \;, \label{lphi} 
\end{equation}
with 
\begin{equation}
\label{act0}
S_{0}^{\phi}  =  \int d^4x\; \left(  \frac{1}{2}  (D^{ab}_{\mu}\phi^{b})^{2}  + \frac{m^2_{\phi}}{2} \phi^a \phi^a
+  \frac{\lambda}{4!}(\phi^{a}\phi^{a})^{2}
+ \tilde{\eta}^{ac}(\partial_{\mu} D^{ab}_{\mu})\eta^{bc} 
- \tilde{\theta}^{ac}(\partial_{\mu} D^{ab}_{\mu})\theta^{bc}      - gf^{abc}(\partial_{\mu}\tilde{\theta}^{ae})(D^{bd}_{\mu}c^{d})\eta^{ce}  \right)  \;  
\end{equation}
and
\begin{equation}
S_{\sigma}  =  \sigma^{2}g  \int d^4x   \; f^{abc}\phi^{a}(\eta^{bc} + \tilde{\eta}^{bc}) \;.     \label{ss}
\end{equation}
As in the case of the Gribov-Zwanziger action, the auxiliary fields $(\tilde{\eta}^{ab},\eta^{ab})$, $(\tilde{\theta}^{ab},\theta^{ab})$ appear quadratically, so that they can be easily integrated out, giving back precisely the non-local starting expression \eqref{acs}. Moreover, in full analogy with the Gribov-Zwanziger case, the local action $S_{loc}^{\phi}$ exhibits a soft breaking of the BRST symmetry. In fact, making use of eqs.\eqref{brst1} and of 
\begin{eqnarray}
&&
s\phi^{a}=-gf^{abc}\phi^{b}c^{c} \;,    \nonumber \\
&&
s\tilde{\theta}^{ab} = \tilde{\eta}^{ab}\;, \qquad s\tilde{\eta}^{ab} =0\;, \nonumber \\
&&
s\eta^{ab}=\theta^{ab}\;, \qquad s\theta^{ab}=0\;, 
\end{eqnarray}
it follows that 
\begin{equation}
s  S_{loc}^{\phi} = \sigma^2 \Delta^{\phi}   \;, \label{bs}
\end{equation}
 where 
\begin{equation}
 \Delta^{\phi}  = g \int d^4x\; f^{abc} \left( -g f^{amn} \phi^{m} c^n (\eta^{bc} + \tilde{\eta}^{bc}) + \phi^a \theta^{bc}     \right)   \;. \label{dphi}
\end{equation}
Again, being of dimension two in the fields, the breaking term  $\Delta^{\phi} $ is a soft breaking. \\\\We now add the local action \eqref{lphi} to the Gribov-Zwanziger action \eqref{sgz}, obtaining 
\begin{eqnarray}
\label{actlc}
S_{ loc} &=& \int d^4x\; \Biggl\{
\frac{1}{4}F^a_{\mu\nu} F^a_{\mu\nu}
+ b^{a}\partial_{\mu}A^{a}_{\mu}
+ \bar{c}^{a}\partial_{\mu}D^{ab}_{\mu}c^{b}
+ \frac{1}{2}(D^{ab}_{\mu}\phi^{b})^{2}  + \frac{m^2_{\phi}}{2} \phi^a \phi^a
+ \frac{\lambda}{4!}(\phi^{a}\phi^{a})^{2}
+ \varphi^{ac}_{\nu}\partial_{\mu}D^{ab}_{\mu}\bar{\varphi}^{bc}_{\nu}
\nonumber \\
&&
- \omega^{ac}_{\nu}\partial_{\mu}D^{ab}_{\mu}\bar{\omega}^{ac}_{\nu}
+ \gamma^{2}gf^{abc}A^{a}_{\mu}(\varphi^{bc}_{\mu} + \bar{\varphi}^{bc}_{\mu})
- gf^{abc}(\partial_{\mu}\bar{\omega}^{ae}_{\nu})(D^{bd}_{\mu}c^{d})\varphi^{ce}_\nu
- \gamma^{4}4(N^{2}-1)
\nonumber \\
&&
+ \tilde{\eta}^{ac}(\partial_{\mu} D^{ab}_{\mu})\eta^{bc}
- \tilde{\theta}^{ac}(\partial_{\mu} D^{ab}_{\mu})\theta^{bc}
+ \sigma^{2}gf^{abc}\phi^{a}(\eta^{bc} + \tilde{\eta}^{bc})
- gf^{abc}(\partial_{\mu}\tilde{\theta}^{ae})(D^{bd}_{\mu}c^{d})\eta^{ce}
\Biggr\}\;.
\end{eqnarray}
As it happens in the case of the Gribov-Zwanziger action, the local action  $S_{ loc}$ can be proven to be renormalizable to all orders. This important property follows from the existence of a large set of Ward identities which can be derived in the matter scalar sector and which restrict very much the possible allowed counterterms.  For the sake of clarity, the whole Appendix A has been devoted to the detailed algebraic proof of the renormalizability of the action \eqref{actlc}. \\\\As in the case of the Gribov-Zwanziger action, expression \eqref{actlc} is well suited to investigate the correlation function 
\begin{equation} 
 \langle {\cal R}^{ab}(x)  {\cal R}^{cd}(y)  \rangle    \;, \label{cphi} 
\end{equation} 
\begin{equation}
{\cal R}^{ab}(x)  =  g \int d^4z\;  ({\cal M}^{-1})^{ac} (x,z)   \; f^{cdb} \phi^{d}(z)   \label{rmsc}  \;, 
\end{equation}
and its relation with the soft BRST breaking in the scalar field sector, eq.\eqref{bs}. In fact, repeating the same reasoning of eqs.\eqref{ss}, \eqref{loce},\eqref{brstbr}, one is led to consider the exact BRST correlation function in the matter scalar field sector
\begin{equation} 
\langle \; s ( \eta^{ab}(x) {\tilde \theta}^{cd}(y)  \; ) \rangle_{S_{ loc}}  = \langle   \theta^{ab}(x) {\tilde \theta}^{cd}(y) + \eta^{ab}(x) {\tilde \eta}^{cd}(y)      \rangle_{S_{ loc}} \;.
\end{equation}
Integrating out the auxiliary fields $(\tilde{\theta}^{ab}, \theta^{ab}, \tilde{\eta}^{ab},\eta^{ab})$ in expression 
\begin{equation}
\int [{\cal D} {\Phi}] \; \left( \theta^{ab}(x) {\tilde \theta}^{cd}(y) + \eta^{ab}(x) {\tilde \eta}^{cd}(y) \right)   \; e^{-S_{loc}} \;,  \label{locphi}
\end{equation}
gives
\begin{equation} 
\frac{ \int [{\cal D} {\Phi}] \;   \left( s \left( \eta^{ab}(x) {\tilde \theta}^{cd}(y)  \right)   \right) \; e^{-S_{loc}} }{ \int [{\cal D} {\Phi}]    \; e^{-S_{loc}}}  =\sigma^4 \; \frac{  \int {\cal D}A {\cal D}{\phi}\; \delta(\partial A) \; \left( det{\cal M} \right) \; {\cal R}^{ab}(x)  {\cal R}^{cd}(y)  \; e^{-(S_{YM}+\gamma^4 H(A) + S^{\phi})} } {\int {\cal D}A {\cal D}{\phi} \; \delta(\partial A) \; \left( det{\cal M} \right)   \; e^{-(S_{YM}+\gamma^4 H(A) + S^{\phi})}  } \;,\label{brstphi}
\end{equation}
showing that, in analogy with the case of the gauge field,  the correlation function \eqref{cphi}  with a cutoff at the Gribov horizon is directly related to the existence of the BRST breaking in the matter sector. \\\\We can now have a look at the two-point correlation function of the scalar field. Nevertheless, before that, an additional effect has to be taken into account. In very strict analogy with the case of the Refined Gribov-Zwanziger action, eq.\eqref{rgz}, the soft breaking of the BRST symmetry occurring in the scalar matter sector, eq.\eqref{bs}, implies the existence of a non-vanishing BRST exact dimension two condensate, namely 
\begin{equation}
\langle s ( \tilde{\theta}^{ab}(x)  {\eta}^{ab}(x) ) \rangle = \langle ( \tilde{\eta}^{ab}(x)  {\eta}^{ab}(x)  -  \tilde{\theta}^{ab}(x)  {\theta}^{ab}(x) ) \rangle \neq 0 \;.  \label{condphi}
\end{equation}
In order to show that expression \eqref{condphi} in non-vanishing, we couple the operator $( \tilde{\eta}^{ab}(x)  {\eta}^{ab}(x)  -  \tilde{\theta}^{ab}(x)  {\theta}^{ab}(x) ) $ to the local action $S_{ loc}$, eq.\eqref{actlc}, by means of a constant external source $J$, 
\begin{equation}
S_{ loc} - J \int d^4x\;  ( \tilde{\eta}^{ab}(x)  {\eta}^{ab}(x)  -  \tilde{\theta}^{ab}(x)  {\theta}^{ab}(x) )  \;, \label{cj}
\end{equation}
and we evaluate the vacuum energy $\mathcal{E}(J)$ in the presence of $J$, namely 
\begin{equation}
 e^{-V\mathcal{E}(J)}= \int {\cal D}{\Phi} \; e^{ -\left( S_{ loc} - J \int d^4x\;  ( \tilde{\eta}^{ab}(x)  {\eta}^{ab}(x)  -  \tilde{\theta}^{ab}(x)  {\theta}^{ab}(x) ) \right) }   \;. \label{ej}
\end{equation}
Thus, the condensate $\langle ( \tilde{\eta}^{ab}(x)  {\eta}^{ab}(x)  -  \tilde{\theta}^{ab}(x)  {\theta}^{ab}(x) ) \rangle$  is obtained by differentiating $\mathcal{E}(J)$ with respect to $J$ and setting $J=0$ at the end, {\it i.e.}
\begin{equation}
\frac{\partial \mathcal{E}(J)}{\partial J} \Big|_{J=0} = - \langle ( \tilde{\eta}^{ab}(x)  {\eta}^{ab}(x)  -  \tilde{\theta}^{ab}(x)  {\theta}^{ab}(x) ) \rangle    \;. \label{vj}
\end{equation}
Employing dimensional regularisation, to the first order, we have 
\begin{equation}
\mathcal{E}(J) = \frac{(N^2-1)}{2} \int \frac{ d^dk}{(2\pi)^d} \; \log\left( k^2 +m^2_{\phi} +\frac{2N\sigma^4 g^2}{k^2+J} \right)  \; + \;{\hat {\cal E}}   \;, \label{fo}
\end{equation}
where ${\hat {\cal E}} $ stands for the part of the vacuum energy which is independent from $J$. Differentiating eq.\eqref{fo} with respect to $J$ and setting $J=0$, we get 
\begin{equation}
 \langle ( \tilde{\eta}^{ab}(x)  {\eta}^{ab}(x)  -  \tilde{\theta}^{ab}(x)  {\theta}^{ab}(x) ) \rangle = (N^2-1) N \sigma^4 g^2  \int \frac{ d^dk}{(2\pi)^d} \frac{1}{k^2} 
 \frac{1}{k^4 + m^2_\phi \;k^2 +  2 N \sigma^4 g^2}   \neq 0 \;. \label{vcondphi}
\end{equation}
Notice that the integral in the right hand side of eq.\eqref{vcondphi} is ultraviolet convergent in $d=4$. Expression  \eqref{vcondphi} shows that, as long as the parameter $\sigma$ in non-vanishing, the condensate $\langle ( \tilde{\eta}^{ab}(x)  {\eta}^{ab}(x)  -  \tilde{\theta}^{ab}(x)  {\theta}^{ab}(x) ) \rangle$ is dynamically generated. \\\\The effect of the condensate 
 \eqref{condphi}  can be taken into account by adding to the action $S_{ loc}$ the novel term 
\begin{equation}
\mu^2_\phi \int d^4x \; s ( \tilde{\theta}^{ab} {\eta}^{ab} )  = \mu^2_\phi \int d^4x\;  ( \tilde{\eta}^{ab}  {\eta}^{ab}  -  \tilde{\theta}^{ab}  {\theta}^{ab} )   \;,  \label{accondphi}
\end{equation}
giving rise to the  Refined action 
\begin{equation}
{\tilde S}_{Ref} =  S_{ loc} +  \int d^4x \left(  \frac{m^2}{2} A^a_\mu A^a_\mu  - \mu^2 \left( {\bar \varphi}^{ab}_{\mu}  { \varphi}^{ab}_{\mu} -  {\bar \omega}^{ab}_{\mu}  { \omega}^{ab}_{\mu} \right)   \right)    - \mu^2_\phi \int d^4x\;  \left( \tilde{\eta}^{ab}  {\eta}^{ab}  -  \tilde{\theta}^{ab}  {\theta}^{ab} \right)  \;. \label{refphi}
\end{equation}
Finally, for the propagator of the scalar field, we get 
\begin{equation} 
\langle  \phi^a(k)  \phi^b(-k) \rangle   =   \delta^{ab}   \frac{k^2 +\mu^2_{\phi}}{k^4 + (\mu^2_\phi+m^2_{\phi})k^2 + 2Ng^2\sigma^4 + \mu^2_\phi m^2_\phi}  \;. \label{phiprop}
\end{equation} 
which is precisely of the the same kind employed in the previous section in order to fit the lattice data.  

\subsection{The quark field} 
In this subsection we generalise the previous construction to the case of quark fields. The  starting non-local action \eqref{acs} is now given by 
\begin{equation}
\label{apsi}
S^{\psi} = \int d^4x\; \left( {\bar \psi}^{i} \gamma_\mu D_{\mu}^{ij} \psi^{j} - m_{\psi}  {\bar \psi}^{i} \psi^{i}  \right) 
 - M^3 g^2   \int d^{4}x\;d^{4}y\;   {\bar \psi}^{i}(x)  (T^a)^{ij} \left[ {\cal M}^{-1}\right]^{ab} (x,y)  (T^b)^{jk} \psi^{k}(y) \;, 
\end{equation}
where the massive parameter $M$ is the analogue of the parameter $\sigma$ of the scalar field and 
\begin{equation}
D^{ij}_\mu = \delta^{ij} \partial_\mu - i g  (T^a)^{ij} A^a_\mu \;, \label{covf}
\end{equation}
is the covariant derivative in the fundamental representation, specified by the generators $(T^a)^{ij}$. As in the previous case, the non-local action \eqref{apsi} can be cast in local form through the introduction of a suitable set of auxiliary fields: $({\bar \lambda}^{ai}, {\lambda}^{ai})$ and $({\bar \xi}^{ai}, {\xi}^{ai})$. The fields $({\bar \lambda}^{ai}, {\lambda}^{ai})$ are Dirac spinors  with two color indices $(a,i)$ belonging, respectively,  to the adjoint and to the  fundamental representation. Similarly, $({\bar \xi}^{ai}, {\xi}^{ai})$ are a pair of spinor fields with ghost number $(-1,1)$. The spinors  $({\bar \lambda}^{ai}, {\lambda}^{ai})$ are anti-commuting, while $({\bar \xi}^{ai}, {\xi}^{ai})$ are commuting. \\\\For the local version of the action, we get 
\begin{equation}
S^{\psi}_{loc} = S_0 + S_M \;, \label{locpsi}
\end{equation}
where 
\begin{equation}
S_0 =   \int d^4x\; \left( {\bar \psi}^{i} \gamma_\mu D_{\mu}^{ij} \psi^{j} - m_{\psi}  {\bar \psi}^{i} \psi^{i}   +  {\bar \lambda}^{ai}( -\partial_\mu D^{ab}_\mu) \lambda^{bi} 
+ {\bar \xi}^{ai}( -\partial_\mu D^{ab}_{\mu} ) \xi^{bi}   - (\partial_\mu {\bar \xi}^{ai}) g f^{acb} (D^{cm}_\mu c^m) \lambda^{bi}  \right)  \;, \label{szpsi}
\end{equation}
and 
\begin{equation}
S_M = g M^{3/2} \int d^4x \; \left(   {\bar \lambda}^{ai} (T^a)^{ij} \psi^{j} +  {\bar \psi}^{i} (T^a)^{ij} \lambda^{aj}    \right)    \;. \label{sM}
\end{equation}
The non-local action $S^{\psi}$ is easily recovered by integrating out the auxiliary fields $({\bar \lambda}^{ai}, {\lambda}^{ai})$ and $({\bar \xi}^{ai}, {\xi}^{ai})$. As in the case of the scalar field, the term $S_M$ induces a soft breaking of the BRST symmetry. In fact, from 
\begin{eqnarray}
s \psi^{i} & = & -ig c^a (T^a)^{ij} \psi^{j} \;, \nonumber \\
s {\bar \psi}^{i} & = & -ig {\bar \psi}^{j} c^a (T^a)^{ji} \;, \nonumber \\
s {\bar \xi}^{ai} & = & {\bar \lambda}^{ai} \;, \qquad s {\bar \lambda}^{ai} = 0\;, \nonumber \\
s {\lambda}^{ai} & = & {\xi}^{ai} \;, \qquad s{\xi}^{ai}=0 \;,  \label{spsi}
\end{eqnarray}
one easily checks that 
\begin{equation}
s S^{\psi}_{loc} = s  S_M = M^{3/2}  \Delta^M   \;, \label{sbM} 
\end{equation}
where
\begin{equation}
\Delta^M = \int d^4x \; \left(  ig^2  {\bar \lambda}^{ai} (T^a)^{ij} c^b (T^b)^{jk}\psi^{k} -ig^2 {\bar \psi}^{k} c^b (T^b)^{ki}(T^a)^{ij} \lambda^{aj}  
- g {\bar \psi}^{i} (T^a)^{ij} \xi^{aj}   \right)   \;. \label{dm}
\end{equation}
Again, being of dimension $5/2$ in the fields, $\Delta^M$ is a soft breaking. In the present case, for the quantity  \eqref{rmatter} we have 
\begin{eqnarray}
{\cal R}^{ai}_{\alpha} (x)  & = &  g \int d^4z\;  ({\cal M}^{-1})^{ab} (x,z)   \;(T^b)^{ij} \psi^{j}_{\alpha} (z)     \;, \nonumber \\
{\bar {\cal R}}^{bj}_{\beta} (x)  & = &  g \int d^4z\;  ({\cal M}^{-1})^{bc } (x,z)  {\bar \psi}^{k}_{\beta}(z)  \;(T^c)^{kj}     \;, \label{rpsi}
\end{eqnarray}
where we have explicitated  the Dirac indices $\alpha,\beta=1,2,3,4$. \\\\As in the case of the scalar field, the action $S^{\psi}_{loc} $ can be added to the Gribov-Zwanziger action. The resulting action, $(S_{GZ} + S^{\psi}_{loc})$, turns out to be renormalizable. Although we shall not give here the details of the proof of the renormalizability of the action $(S_{GZ} + S^{\psi}_{loc})$, it is worth mentioning that it can be given by following the framework already outlined in \cite{Baulieu:2009xr}, where a similar non-local spinor action has been considered. \\\\Proceeding now as in the case of the scalar field, one finds 
\begin{equation} 
\frac{ \int [{\cal D} {\Phi}] \;   \left( s \left( {\bar \xi}^{ai}_{\alpha}(x) {\lambda}^{bj}_\beta(y)  \right)   \right) \; e^{-(S_{GZ}+S^{\psi}_{loc})} }{ \int [{\cal D} {\Phi}]    \; e^{-(S_{GZ}+S^{\psi}_{loc})}}  =M^3 \; \frac{  \int {\cal D}A {\cal D}{\psi} {\cal D}{\bar \psi} \; \delta(\partial A)  \left( det{\cal M} \right) {\cal R}^{ai}_{\alpha}(x)  {\bar {\cal R}}^{bj}_{\beta}(y)  \; e^{-(S_{YM}+\gamma^4 H(A) + S^{\psi})} } {\int {\cal D}A {\cal D}{\psi} {\cal D}{\bar \psi} \; \delta(\partial A) \; \left( det{\cal M} \right)   \; e^{-(S_{YM}+\gamma^4 H(A) + S^{\psi})}  } \;,\label{brstpsi}
\end{equation}
showing that  the correlation function $\langle {\cal R}^{ai}_{\alpha}(x)  {\bar {\cal R}}^{bj}_{\beta}(y) \rangle$  with a cutoff at the Gribov horizon is  related to the existence of the BRST breaking, eq.\eqref{sbM}. \\\\Let us end this section by discussing the two-point correlation function of the quark field. As before, an additional effect has to be taken into account. Also here,  the soft breaking of the BRST symmetry, eq.\eqref{sbM}, implies the existence of a non-vanishing BRST exact dimension two condensate, namely 
\begin{equation}
\langle s ( {\bar {\xi}}^{ai}(x)  {\lambda}^{ai}(x) ) \rangle = \langle ( {\bar \lambda}^{ai}(x)  {\lambda}^{ai}(x)  + {\bar  \xi}^{ai}(x)  {\xi}^{ai}(x) ) \rangle \neq 0 \;,  \label{condpsi}
\end{equation}
whose effect can be taken into account by adding to the action $S^{\psi}_{loc}$ the term 
\begin{equation}
\mu^2_\psi \int d^4x \; s   ( {\bar {\xi}}^{ai}(x)  {\lambda}^{ai}(x) )   = \mu^2_\psi \int d^4x\;  ( {\bar \lambda}^{ai}(x)  {\lambda}^{ai}(x)  + {\bar  \xi}^{ai}(x)  {\xi}^{ai}(x) )  \;.  \label{accondpsi}
\end{equation}
Therefore, including the dimension two condensates, we end up with the Refined action 
\begin{equation}
{\tilde S}_{Ref}^{\psi} =  S_{RGZ} + S^{\psi}_{loc} +  \mu^2_\psi \int d^4x\;  ( {\bar \lambda}^{ai}(x)  {\lambda}^{ai}(x)  + {\bar  \xi}^{ai}(x)  {\xi}^{ai}(x) )   \;. \label{refpsi}
\end{equation}
Finally, for the propagator of the quark field, we get 
\begin{equation} 
\langle  \psi^{i}(k)  {\bar \psi}^{j}(-k) \rangle   =   \delta^{ij} \;  \frac{-ik_\mu \gamma_\mu + {\cal A}(k^2)}{k^2 + {\cal A}^2(k^2)}  \;, \label{psiprop}
\end{equation} 
where 
\begin{equation}
{\cal A}(k^2) = m_{\psi} + \frac{g^2 M^3 C_F}{k^2+\mu^2_\psi} \;, \label{A}
\end{equation}
and 
\begin{equation}
 (T^a)^{ij} (T^a)^{jk} = \delta^{ik} C_F \;, \qquad C_F= \frac{N^2-1}{2N}  \;.    \label{norm}
\end{equation}
Expression \eqref{psiprop} is of the the same kind employed  to fit the lattice data.

\section{Conclusion}

One of the striking features of the (R)GZ formulation of non-perturbative Euclidean continuum Yang-Mills theories is the appearance of the soft breaking of the BRST symmetry, which seems deeply related to gluon confinement. Recently, direct lattice investigations \cite{Cucchieri:2014via} have confirmed the existence of this breaking through the analysis of the correlation function:
\begin{eqnarray}
\langle \tilde {\cal R}^{ab}_{\mu}(k)\tilde {\cal R}^{cd}_{\nu}(-k)\rangle&\stackrel{k\to 0}{\sim}&\frac{1}{k^4} \label{RRgluon}
\\
{\cal R}^{ac}_{\mu}(x)&=&g \int d^4z ({\cal M}^{-1})^{ad}(x,z) f^{dec}A^e_{\mu}(z)
\,, 
\end{eqnarray}
As pointed in \cite{Cucchieri:2014via}, this non-vanishing correlator  signals the breaking of  the BRST invariance.
Interestingly enough, the behaviour \eqref{RRgluon} is in quite good agreement with the RGZ framework.

The aim of the present work is that of  providing evidence that a similar picture can be consistently achieved in the matter sector. The cases of both adjoint scalars and quarks indicate that it is possible to introduce an analogous operator ${\cal R}^{ai}_{F}$ for matter fields,
\begin{eqnarray}
{\cal R}^{ai}_F(x)  &=&  g \int d^4z\;  ({\cal M}^{-1})^{ab} (x,z)   \;(T^b)^{ij} \;F^{j}(z)  
\,,
\end{eqnarray}
so that the correlation function $\langle{\cal R}_F{\cal R}_F\rangle$ is non-vanishing and, from the available lattice data, seems to behave like expression \eqref{RRgluon}, namely
\begin{eqnarray}
\langle \tilde {\cal R}^{ai}_F(k)\tilde {\cal R}^{bj}_F(-k)\rangle&\stackrel{k\to 0}{\sim}&\frac{1}{k^4} 
\,.
\end{eqnarray}
Again, the non-vanishing of $\langle{\cal R}_F{\cal R}_F\rangle$ indicates the soft breaking of the BRST symmetry in the matter sector.
In this sense, the correlation function $\langle{\cal R}_F{\cal R}_F\rangle$ could be regarded as a direct signature for BRST breaking, being accessible both analytically as well as through numerical lattice simulations. 

Concerning the analytic side, we have been able to construct a local and renormalizable action including matter fields which accommodates the non-trivial correlation functions $\langle{\cal R}_F{\cal R}_F\rangle$. Our analysis further suggests that the inverse of the Faddeev-Popov operator ${\cal M}^{-1}$, whose existence is guaranteed by the restriction to the Gribov region $\Omega$, couples in a universal way to any colored field $G^i$ ({\it e.g.} gluon and matter fields),
\begin{eqnarray}
{\cal R}^{ai}_G(x)  &=&  g \int d^4z\;  ({\cal M}^{-1})^{ab} (x,z)   \;(T^b)^{ij} \;G^{j}(z)  \label{RG}
\,,
\end{eqnarray}giving rise to a non-vanishing correlation function 
\begin{eqnarray}
\langle \tilde {\cal R}_G(k)\tilde {\cal R}_G(-k)\rangle&\stackrel{k\to 0}{\sim}&\frac{1}{k^4} \label{RRG}
\,.
\end{eqnarray}

Therefore, these correlation functions could signal that the soft breaking of the BRST invariance generated by the restriction to the Gribov region is transmitted to the colored objects through the coupling with the inverse Faddeev-Popov operator $({\cal M}^{-1})^{ab}$, as described by equations \eqref{RG} and \eqref{RRG}.

Although this construction has been presented in the case of the Landau gauge, it can be generalized to other gauges, like, {\it e.g.}, the Maximal Abelian Gauge \cite{MAG-wip}.

\section*{Acknowledgments}
The authors would like to thank D. Dudal and O. Oliveira for useful discussions and help with Fig.3. We thank A. Maas for discussions and also for providing the preliminary lattice data on the adjoint scalars.

The Conselho Nacional de Desenvolvimento Cient\'{\i}fico e
Tecnol\'{o}gico (CNPq-Brazil), the Faperj, Funda{\c{c}}{\~{a}}o de
Amparo {\`{a}} Pesquisa do Estado do Rio de Janeiro,  the
Coordena{\c{c}}{\~{a}}o de Aperfei{\c{c}}oamento de Pessoal de
N{\'{\i}}vel Superior (CAPES)  are gratefully acknowledged. L.F.P. is supported by a BJT fellowship from the brazilian program ``Ci\^encia sem Fronteiras'' (grant number 301111/2014-6).

\begin{appendix}
\section{Algebraic Renormalization of the scalar action $S_{ loc}$}
In order to prove the renormalizability of the action $S_{ loc}$, eq.\eqref{actlc}, we proceed as in   \cite{Zwanziger:1988jt,Zwanziger:1989mf,Zwanziger:1992qr,Dudal:2007cw,Dudal:2008sp,Dudal:2011gd} and we embed the theory into an extended action $\Sigma$ enjoying exact BRST symmetry, given by
\begin{eqnarray}
\label{fullact}
\Sigma &=& \int d^4x\; \Biggl\{
\frac{1}{4}F^a_{\mu\nu} F^a_{\mu\nu}
+ b^{a}\partial_{\mu}A^{a}_{\mu}
+ \bar{c}^{a}\partial_{\mu}D^{ab}_{\mu}c^{b}
+ \frac{1}{2}(D^{ab}_{\mu}\phi^{b})^{2} + \frac{m^2_{\phi}}{2} \phi^a \phi^a
+ \frac{\lambda}{4!}(\phi^{a}\phi^{a})^{2}
+ \bar{\varphi}^{ac}_{\nu}\partial_{\mu}D^{ab}_{\mu}\varphi^{bc}_{\nu}
\nonumber \\
&&
- \bar{\omega}^{ac}_{\nu}\partial_{\mu}D^{ab}_{\mu}\omega^{bc}_{\nu}
- gf^{abc}(\partial_{\mu}\bar{\omega}^{ae}_{\nu})(D^{bd}_{\mu}c^{d})\varphi^{ce}_\nu
-{N}^{ac}_{\mu\nu}\,D^{ab}_{\mu}\bar{\omega}^{bc}_{\nu}
-{M}^{ae}_{\mu\nu}\Bigl[D^{ab}_{\mu}\bar{\varphi}^{be}_{\nu}
-gf^{abc}(D^{bd}_{\mu}c^{d})\bar{\omega}^{ce}_{\nu}
\Bigr]
\nonumber \\
&&
- \bar{M}^{ac}_{\mu\nu}\,D^{ab}_{\mu}\varphi^{bc}_{\nu}
+ \bar{N}^{ae}_{\mu\nu}\Bigl[D^{ab}_{\mu}\omega^{be}_{\nu}
- gf^{abc}(D^{bd}_{\mu}c^{d})\varphi^{ce}_{\nu}
\Bigr]
- \bar{M}^{ac}_{\mu\nu}{M}^{ac}_{\mu\nu}
+ \bar{N}^{ac}_{\mu\nu}{N}^{ac}_{\mu\nu}
+ \tilde{\eta}^{ac}(\partial_{\mu}D_{\mu}^{ab})\eta^{bc}
\nonumber \\
&&
- \tilde{\theta}^{ac}(\partial_{\mu}D_{\mu}^{ab})\theta^{bc}
- gf^{abc}(\partial_{\mu}\tilde{\theta}^{ae})(D^{bd}_{\mu}c^{d})\eta^{ce}
+ gf^{abc}\tilde{V}^{ad}\phi^{b}\eta^{cd}
+ gf^{abc}V^{ad}\left( 
- gf^{bde}\phi^{d}c^{e}\tilde{\theta}^{cd}
+ \phi^{b}\tilde{\eta}^{cd}
\right)
\nonumber \\
&&
+ \rho\left(\tilde{V}^{ab}V^{ab} - \tilde{U}^{ab}U^{ab}\right)
+ gf^{abc}\tilde{U}^{al}\left( gf^{bde}\phi^{d}c^{e}\eta^{cl}
- \phi^{b}\theta^{cl}\right)
+ gf^{abc}U^{ad}\phi^{b}\tilde{\theta}^{cd}
- K^{a}_{\mu}D^{ab}_{\mu}c^{b}
+ \frac{g}{2}f^{abc} L^{a}c^{b}c^{c}
\nonumber \\
&&
- gf^{abc} F^{a}\phi^{b}c^{c}
\Biggr\}\;,
\end{eqnarray}
where $\left( M^{ab}_{\mu\nu}, \bar{M}^{ab}_{\mu\nu}, N^{ab}_{\mu\nu}, \bar{N}^{ab}_{\mu\nu}, V^{abc}, \tilde{V}^{abc}, U^{abc}, \tilde{U}^{abc} \right)$ are external sources. 
\noindent The original local action $S_{ loc}$, \eqref{actlc}, can be re-obtained from the extended action $\Sigma$  by letting the external fields to assume their  physical values namely
\begin{eqnarray}
\label{physval1}
&&
M^{ab}_{\mu\nu}\Big{|}_{phys}=\bar{M}^{ab}_{\mu\nu}\Big{|}_{phys}=
\gamma^{2}\delta^{ab}\delta_{\mu\nu}
\;;\nonumber \\
&&
V^{ab}\Big{|}_{phys}=\tilde{V}^{ab}\Big{|}_{phys}=
\sigma^{2}\delta^{ab}
\;;\nonumber \\
&&
N^{ab}_{\mu\nu}\Big{|}_{phys}=\bar{N}^{ab}_{\mu\nu}\Big{|}_{phys}=
U^{ab}\Big{|}_{phys}=\tilde{U}^{ab}\Big{|}_{phys}=0
\;. \nonumber\\
&& K^{a}_\mu = L^a = F^a =0 \;, 
\end{eqnarray}
so that 
\begin{equation} 
\Sigma \Big|_{phys}  = S_{ loc}  + V \rho \;\sigma^4 g^2 N(N^2-1)  \;, \label{limit} 
\end{equation}
where the parameter $\rho$ has been introduced in order to take into account possible divergences in the vacuum energy associated to the term $\sigma^4$. This term stems from the source term $\rho\tilde{V}^{abc}V^{abc}$, which  is allowed by power counting. In the physical limit the vertex $\phi c \tilde{\theta}$ remains non-vanishing. Though, it is harmless, due to the absence of mixed propagators $\langle c\; \tilde{\theta}\rangle $ and $\langle {\bar c}\; \theta\rangle $. \\\\It is easy to check that the extended action $\Sigma$ enjoys exact BRST invariance, {\it i.e.} 
\begin{equation} 
s \Sigma = 0  \;, \label{sinv}
\end{equation}
where
\begin{eqnarray}
\label{brst1}
sA^{a}_{\mu} &=& - D^{ab}_{\mu}c^{b}\;,\nonumber \\
s\phi^{a}& = & -gf^{abc}\phi^{b}c^{c} \;,    \nonumber \\
s c^{a} &=& \frac{1}{2}gf^{abc}c^{b}c^{c} \;, \nonumber \\
s{\bar c}^{a} &=& b^{a}\;, \qquad \; \; 
sb^{a} = 0 \;, \nonumber \\
s{\bar \omega}^{ab}_\mu & = & {\bar \varphi}^{ab}_\mu \;, \qquad  s {\bar \varphi}^{ab}_\mu =0\;, \nonumber \\
s { \varphi}^{ab}_\mu&  = & {\omega}^{ab}_\mu  \;, \qquad s {\omega}^{ab}_\mu = 0 \;,  \nonumber \\
s\tilde{\theta}^{ab} & =&  \tilde{\eta}^{ab}\;, \qquad s\tilde{\eta}^{ab} =0\;, \nonumber \\
s\eta^{ab}& = & \theta^{ab}\;, \qquad s\theta^{ab}=0\;, 
\end{eqnarray}
and 
\begin{eqnarray}
&&
sM^{ab}_{\mu\nu} = N^{ab}_{\mu\nu}\;; \qquad sN^{ab}_{\mu\nu} = 0\;; \nonumber \\
&&
s\bar{N}^{ab}_{\mu\nu} = \bar{M}^{ab}_{\mu\nu}\;; \qquad s\bar{M}^{ab}_{\mu\nu} = 0\;; \nonumber \\
&&
s\tilde{U}^{ab}=\tilde{V}^{ab}\,,\qquad s\tilde{V}^{ab}=0\;; \nonumber \\
&&
sV^{ab}=U^{ab}\,,\qquad sU^{ab}=0\;; \nonumber \\
&&
sK^{a}=sL^{a}=sF^{a}=0\;.
\end{eqnarray}
As noticed in \cite{Zwanziger:1988jt,Zwanziger:1989mf,Zwanziger:1992qr,Dudal:2007cw,Dudal:2008sp,Dudal:2011gd}, it is useful introducing a multi-index notation for the localizing auxiliary fields $({\bar \varphi}^{ab}_\mu, { \varphi}^{ab}_\mu, {\bar \omega}^{ab}_\mu, {\bar \omega}^{ab}_\mu) = ({\bar \varphi}^{a}_i, { \varphi}^{a}_i, {\bar \omega}^{a}_i, {\bar \omega}^{a}_i) $ where the multi-index $i=(b,\mu)$ runs from $1$ to $4(N^2-1)$. The important reason in order to introduce the multi-index notation is related to the existence of a global symmetry $U(4(N^2-1))$ in the index $i$, which plays an important role in the proof of the algebraic renormalization. Analogously, one can introduce a second index $I$ for the localizing fields of the matter scalar sector $(\tilde{\eta}^{ab},{\eta}^{ab},\tilde{\theta}^{ab},{\theta}^{ab})= (\tilde{\eta}^{aI},{\eta}^{aI},\tilde{\theta}^{aI},{\theta}^{aI})$, where $I=1,..,(N^2-1)$. Again, the introduction of the index $I$ is related to the existence of a second global symmetry $U(N^2-1)$. In the multi-index notation, the action \eqref{fullact} reads 
\begin{eqnarray}
\label{fullactm}
\Sigma &=& \int d^4x\; \Biggl\{
\frac{1}{4}F^a_{\mu\nu} F^a_{\mu\nu}
+ b^{a}\partial_{\mu}A^{a}_{\mu}
+ \bar{c}^{a}\partial_{\mu}D^{ab}_{\mu}c^{b}
+ \frac{1}{2}(D^{ab}_{\mu}\phi^{b})^{2} + \frac{m^2_{\phi}}{2} \phi^a \phi^a 
+ \frac{\lambda}{4!}(\phi^{a}\phi^{a})^{2}
+ \bar{\varphi}^{a}_{i}\partial_{\mu}D^{ab}_{\mu}\varphi^{b}_{i}
\nonumber \\
&&
- \bar{\omega}^{a}_{i}\partial_{\mu}D^{ab}_{\mu}\omega^{b}_{i}
- gf^{abc}(\partial_{\mu}\bar{\omega}^{a}_{i})(D^{bd}_{\mu}c^{d})\varphi^{c}_{i}
-{N}^{a}_{\mu{i}}\,D^{ab}_{\mu}\bar{\omega}^{b}_{i}
-{M}^{a}_{\mu{i}}\Bigl[D^{ab}_{\mu}\bar{\varphi}^{b}_{i}
-gf^{abc}(D^{bd}_{\mu}c^{d})\bar{\omega}^{c}_{i}
\Bigr]
\nonumber \\
&&
- \bar{M}^{a}_{\mu{i}}\,D^{ab}_{\mu}\varphi^{b}_{i}
+ \bar{N}^{a}_{\mu{i}}\Bigl[D^{ab}_{\mu}\omega^{b}_{i}
- gf^{abc}(D^{bd}_{\mu}c^{d})\varphi^{c}_{i}
\Bigr]
- \bar{M}^{a}_{\mu{i}}{M}^{a}_{\mu{i}}
+ \bar{N}^{a}_{\mu{i}}{N}^{a}_{\mu{i}}
+ \tilde{\eta}^{aI}(\partial_{\mu}D_{\mu}^{ab})\eta^{bI}
\nonumber \\
&&
- \tilde{\theta}^{aI}(\partial_{\mu}D_{\mu}^{ab})\theta^{bI}
- gf^{abc}(\partial_{\mu}\tilde{\theta}^{aI})(D^{bd}_{\mu}c^{d})\eta^{cI}
+ gf^{abc}\tilde{V}^{aI}\phi^{b}\eta^{cI}
+ gf^{abc}V^{aI}\left( 
- gf^{bde}\phi^{d}c^{e}\tilde{\theta}^{cI}
+ \phi^{b}\tilde{\eta}^{cI}
\right)
\nonumber \\
&&
+ \rho\left(\tilde{V}^{aI}V^{aI} - \tilde{U}^{aI}U^{aI}\right)
+ gf^{abc}\tilde{U}^{aI}\left( gf^{bde}\phi^{d}c^{e}\eta^{cI}
- \phi^{b}\theta^{cI}\right)
+ gf^{abc}U^{aI}\phi^{b}\tilde{\theta}^{cI}
- K^{a}_{\mu}D^{ab}_{\mu}c^{b} \nonumber \\ &&
+ \frac{g}{2}f^{abc} L^{a}c^{b}c^{c}
- gf^{abc} F^{a}\phi^{b}c^{c}
\Biggr\}\;,
\end{eqnarray}
We are now ready to write down the large set of Ward identities fulfilled by the action \eqref{fullactm}. These are given by: 

\noindent {\bf $\bullet$ The Slavnov-Taylor identity}:
\begin{equation} 
S(\Sigma) = 0  \;, \label{st}
\end{equation} 
where 
\begin{eqnarray}
S(\Sigma) &=& \int d^{4}x \Biggl\{
\frac{\delta \Sigma}{\delta K^{a}_{\mu}}\frac{\delta \Sigma}{\delta A^{a}_{\mu}}
+ \frac{\delta \Sigma}{\delta F^{a}}\frac{\delta \Sigma}{\delta \phi^{a}}
+ \frac{\delta \Sigma}{\delta L^{a}}\frac{\delta \Sigma}{\delta c^{a}}
+ b^{a}\frac{\delta \Sigma}{\delta \bar{c}^{a}}
+ \omega^{a}_{i}\frac{\delta \Sigma}{\delta \varphi^{a}_{i}}
+ \bar{\varphi}^{a}_{i}\frac{\delta \Sigma}{\delta \bar{\omega}^{a}_{i}}
\nonumber \\
&&
+ \tilde{\eta}^{aI}\frac{\delta \Sigma}{\delta \tilde{\theta}^{aI}}
+ \theta^{aI}\frac{\delta \Sigma}{\delta \eta^{aI}}
+ N^{a}_{\mu i}\frac{\delta \Sigma}{\delta M^{a}_{\mu i}}
+ \bar{M}^{a}_{\mu i}\frac{\delta \Sigma}{\delta \bar{N}^{a}_{\mu i}}
+ \tilde{V}^{aI}\frac{\delta \Sigma}{\delta \tilde{U}^{aI}}
+ U^{aI}\frac{\delta \Sigma}{\delta V^{aI}}
\Biggr\}  \;.     \label{stop}
\end{eqnarray}
For future convenience, let us also introduce the so-called linearized Slavnov-Taylor operator ${\cal B}_{\Sigma}$, given by 
\begin{eqnarray}
{\cal B}_{\Sigma} &=& \int d^{4}x \Biggl\{
\frac{\delta \Sigma}{\delta K^{a}_{\mu}}\frac{\delta }{\delta A^{a}_{\mu}}
+ \frac{\delta \Sigma}{\delta A^{a}_{\mu}}\frac{\delta }{\delta K^{a}_{\mu}}
+ \frac{\delta \Sigma}{\delta F^{a}}\frac{\delta }{\delta \phi^{a}}
+ \frac{\delta \Sigma}{\delta \phi^{a}}\frac{\delta }{\delta F^{a}}
+ \frac{\delta \Sigma}{\delta L^{a}}\frac{\delta }{\delta c^{a}}
+ \frac{\delta \Sigma}{\delta c^{a}}\frac{\delta }{\delta L^{a}}
+ b^{a}\frac{\delta }{\delta \bar{c}^{a}}
\nonumber \\
&&
+ \omega^{a}_{i}\frac{\delta }{\delta \varphi^{a}_{i}}
+ \bar{\varphi}^{a}_{i}\frac{\delta }{\delta \bar{\omega}^{a}_{i}}
+ \tilde{\eta}^{aI}\frac{\delta }{\delta \tilde{\theta}^{aI}}
+ \theta^{aI}\frac{\delta }{\delta \eta^{aI}}
+ N^{a}_{\mu i}\frac{\delta }{\delta M^{a}_{\mu i}}
+ \bar{M}^{a}_{\mu i}\frac{\delta }{\delta \bar{N}^{a}_{\mu i}}
+ \tilde{V}^{aI}\frac{\delta }{\delta \tilde{U}^{aI}}
+ U^{aI}\frac{\delta }{\delta V^{aI}}
\Biggr\}\;.  \nonumber \\    \label{lst}
\end{eqnarray}
The operator ${\cal B}_{\Sigma}$ enjoys the important property of being nilpotent
\begin{equation}
{\cal B}_{\Sigma} {\cal B}_{\Sigma} = 0 \;.    \label{nlst}
\end{equation}
{\bf $\bullet$  The gauge-fixing and anti-ghost equations}:
\begin{equation}
\frac{\delta\Sigma}{\delta b^{a}}=\partial_{\mu}A^{a}_{\mu}\,,\qquad
\frac{\delta\Sigma}{\delta\bar{c}^{a}}+\partial_{\mu}\frac{\delta\Sigma}{\delta K^{a}_{\mu}}=0\,.
\label{GFandAntiGhost}
\end{equation}

\noindent {\bf $\bullet$  The linearly broken Ward identities}:
\begin{eqnarray}
&&\frac{\delta \Sigma}{\delta \bar{\varphi}^{a}_{i}} + \partial_{\mu}\frac{\delta\Sigma}{\delta \bar{M}^{a}_{\mu i}}  = 0\,,
\\
&&\frac{\delta\Sigma}{\delta\omega^{a}_{i}} + \partial_{\mu}\frac{\delta\Sigma}{\delta N^{a}_{\mu i}} - gf^{abc} \frac{\delta \Sigma}{\delta b^{c}} \bar{\omega}^{b}_{i} = 0\,,
\\
&&
\frac{\delta \Sigma}{\delta \bar{\omega}^{a}_{i}} 
+ \partial_{\mu}\frac{\delta\Sigma}{\delta\bar{N}^{a}_{\mu i}} 
- gf^{abc} M^{b}_{\mu i}\frac{\delta\Sigma}{\delta K^{c}_{\mu}} = 0\,,
\\
&&
\frac{\delta\Sigma}{\delta\varphi^{a}_{i}} 
+ \partial_{\mu}\frac{\delta\Sigma}{\delta M^{a}_{\mu i}} 
- gf^{abc}\left( \frac{\delta \Sigma}{\delta b^{c}}\bar{\varphi}^{b}_{i} 
+ \frac{\delta \Sigma}{\delta \bar{c}^{b}} \bar{\omega}^{c}_{i} 
- \bar{N}^{c}_{\mu i} \frac{\delta \Sigma}{\delta K^{b}_{\mu}} \right) = 0 \,,
\\
&&
\int d^{4}x\; \left[ c^{a}\frac{\delta }{\delta \omega^{a}_{i}} + \bar{\omega}^{a}_{i}\frac{\delta }{\delta\bar{c}^{a}} + \bar{N}^{a}_{\mu i}\frac{\delta }{\delta K^{a}_{\mu}} \right]\Sigma  = 0  \;,
\\
&&
\int d^{4}x\; \left[ c^{a}\frac{\delta }{\delta \theta^{aI}} + \tilde{\theta}^{aI}\frac{\delta }{\delta\bar{c}^{a}} - \tilde{U}^{aI}\frac{\delta }{\delta F^{a}} \right]\Sigma  = 0  \;,
\\
&&
\int d^{4}x\; \left[ \frac{\delta }{\delta \eta^{bI}} - gf^{abc}\tilde{U}^{aI}\frac{\delta}{\delta F^{c}} - gf^{abe}\left(\tilde{\eta}^{aI}\frac{\delta}{\delta b^{e}}-\tilde{\theta}^{aI}\frac{\delta}{\delta \bar{c}^{e}}\right) \right]\Sigma  = \int d^{4}x\; gf^{abc}V^{aI}\phi^{c}  \;,
\\
&&
\int d^{4}x\; \left[ \frac{\delta }{\delta \theta^{bI}} - gf^{abe}\tilde{\theta}^{aI}\frac{\delta}{\delta b^{e}} \right]\Sigma  = -\int d^{4}x\; gf^{abc}\tilde{U}^{aI}\phi^{c}  \;,
\\
&&
\int d^{4}x\; \left[ \frac{\delta}{\delta \tilde{\theta}^{aI}} - gf^{abc}V^{cI}\frac{\delta}{\delta F^{b}}\right] \Sigma= \int d^{4}x gf^{abc}U^{cI}\phi^{b}\;,
\\
&&
\int d^{4}x\; \frac{\delta \Sigma}{\delta \tilde{\eta}^{bI}} =-  \int d^{4}x\; gf^{abc}V^{aI}\phi^{c}  \;.
\end{eqnarray}

\noindent {\bf $\bullet$ The ghost equation}:
\begin{equation}
{\cal G}^{a}(\Sigma) = \Delta^{a}_{class}\;,
\end{equation}
where
\begin{eqnarray}
{\cal G}^{a} &=& \int d^{4}x \left[ \frac{\delta}{\delta c^{a}} + gf^{abc}\left( \bar{c}^{b}\frac{\delta}{\delta b^{c}} + \bar{\omega}^{b}_{i}\frac{\delta}{\delta \varphi^{c}_{i}} + \varphi^{b}_{i}\frac{\delta}{\delta\omega^{c}_{i}} + M^{b}_{\mu i}\frac{\delta}{\delta N^{c}_{\mu i}} + \bar{N}^{b}_{\mu i}\frac{\delta}{\delta \bar{M}^{c}_{\mu i}} +\tilde{\theta}^{bI}\frac{\delta}{\delta \tilde{\eta}^{cI}}
\right.
\right.
\nonumber \\
&&
\phantom{\int d^{4}x\,}  
\left.
\left.
\eta^{bI}\frac{\delta}{\delta \theta^{cI}} + \tilde{U}^{bI}\frac{\delta}{\delta \tilde{V}^{cI}} + V^{bI}\frac{\delta}{\delta U^{cI}}
\right)\right]
\end{eqnarray}
and
\begin{equation}
\Delta^{a}_{class} = \int d^{4}x gf^{abc}\left( K^{b}_{\mu}A^{c}_{\mu} - L^{b}c^{c} + F^{b}\phi^{c} \right)\;.
\end{equation}

\noindent {\bf $\bullet$ The global symmetry $U(f=4(N^{2}-1))$}:
\begin{eqnarray}
 \mathcal{L}_{ij}(\Sigma) &=&  \int d^{4}x \left[ 
 \varphi^{c}_{i}\frac{\delta}{\delta\varphi^{c}_{j}}
- \bar{\varphi}^{c}_{i}\frac{\delta }{\delta \bar{\varphi}^{c}_{j}}
+ \omega^{c}_{i}\frac{\delta}{\delta\omega^{c}_{j}}    
- \bar{\omega}^{c}_{i}\frac{\delta }{\delta \bar{\omega}^{c}_{j}}
+ M^{c}_{\mu i}\frac{\delta}{\delta{M}^{c}_{\mu j}}  
- \bar{M}^{a}_{\mu i}\frac{\delta }{\delta \bar{M}^{a}_{\mu j}}
 \right. 
\nonumber \\
&&
\left. 
\phantom{\int d^{4}x\,}  
+ N^{a}_{\mu i}\frac{\delta}{\delta{N}^{a}_{\mu j}}  
- \bar{N}^{a}_{\mu i}\frac{\delta }{\delta \bar{N}^{a}_{\mu j}}
\right]\Sigma = 0\;.
\end{eqnarray}

\noindent {\bf $\bullet$  The global symmetry $U(f'=(N^{2}-1))$}:
\begin{eqnarray}
{\cal L}^{IJ}(\Sigma) &=& \int d^{4}x\, \left[
\theta^{bI} \frac{\delta}{\delta \theta^{bJ}} 
- \tilde{\theta}^{bI} \frac{\delta}{\delta \tilde{\theta}^{bJ}}
+ \eta^{bI} \frac{\delta}{\delta \eta^{bJ}}
- \tilde{\eta}^{bI} \frac{\delta}{\delta \tilde{\eta}^{bJ}}
+ V^{aI} \frac{\delta}{\delta V^{aJ}}
- \tilde{V}^{aI} \frac{\delta}{\delta \tilde{V}^{aJ}}
\right.
\nonumber \\
&&
\left.
\phantom{\int d^{4}x\,}  
+ U^{aI} \frac{\delta}{\delta U^{aJ}}
- \tilde{U}^{aI} \frac{\delta}{\delta \tilde{U}^{aJ}}
\right]\Sigma = 0\;.
\end{eqnarray}
Let us also dispslay below the quantum numbers of all fields and sources 

\noindent {\bf $\bullet$ Table of quantum numbers} (``B" is for bosonic fields and ``F" is for fermionic fields) :
\begin{center}
\begin{tabular}{l|c|c|c|c|c|c|c|c|c|c|c|c|c|}
&$A$&$\phi$&$c$&$\bar{c}$&$b$&$\varphi$&$\bar\varphi$&$\omega$&$\bar\omega$&$\eta$&$\tilde{\eta}$&$\theta$&$\tilde{\theta}$\cr
\hline\hline
$\phantom{\Bigl|}\!\!$Dim
&1&$1$&0&2&2&1&1&1&1&1&1&1&1\cr
\hline
$\phantom{\Bigl|}\!\!$Ghost\#
&0&0&1&$-1$&$0$&0&0&1&$-1$&0&0&1&$-1$\cr
\hline
$\phantom{\Bigl|}\!\!$Charge-$q_f$
&0&0&0&0&0&1&$-1$&1&$-1$&0&0&0&0\cr
\hline
$\phantom{\Bigl|}\!\!$Charge-$q_{f'}$
&0&0&0&0&0&0&0&$0$&0&$1$&$-1$&1&$-1$\cr
\hline
$\phantom{\Bigl|}\!\!$Nature
&B&B&F&F&B&B&B&F&F&B&B&F&F
\end{tabular}
\end{center}

\begin{center}
\begin{tabular}{l|c|c|c|c|c|c|c|c|c|c|c|}
&$M$&$\bar{M}$&$N$&$\bar{N}$&$U$&$\tilde{U}$&$V$&$\tilde{V}$&$K$&$L$&$F$\cr
\hline\hline
$\phantom{\Bigl|}\!\!$Dim
&2&2&2&2&2&2&2&2&3&4&3\cr
\hline
$\phantom{\Bigl|}\!\!$Ghost\#
&0&0&1&$-1$&$1$&$-1$&0&0&$-1$&$-2$&$-1$\cr
\hline
$\phantom{\Bigl|}\!\!$Charge-$q_f$
&1&$-1$&1&$-1$&0&0&0&0&0&0&0\cr
\hline
$\phantom{\Bigl|}\!\!$Charge-$q_{f'}$
&0&0&0&0&1&$-1$&1&$-1$&0&$0$&0\cr
\hline
$\phantom{\Bigl|}\!\!$Nature
&B&B&F&F&F&F&B&B&F&B&F
\end{tabular}
\end{center}

\subsection{Algebraic characterisation of the invariant counter term and renormalizability}
In order to determine the most general invariant counterterm which can be freely added to each order of perturbation theory, we follow the Algebraic Renormalization framework  \cite{Piguet:1995er} and perturb  the complete action $\Sigma$ by adding an integrated local polynomial in the fields and sources with dimension bounded by four and vanishing ghost number, $\Sigma_{ct}$, and we require that the perturbed action, $(\Sigma + \varepsilon \Sigma_{ct})$, where $\varepsilon$ is an infinitesimal expansion parameter, obeys the same Ward identities fulfilled by $\Sigma$ to the first order in the parameter $\varepsilon$. Therefore, in the  case of the Slavnov-Taylor identity  \eqref{st}, we have
\begin{equation}
S\left( \Sigma + \varepsilon\Sigma_{ct} \right) = 0 + {\cal O}(\varepsilon^{2})\;,   \label{pertb1}
\end{equation}
which leads to
\begin{equation}
{\cal B}_{\Sigma}\left( \Sigma_{ct} \right) = 0\;, \label{perturblin}
\end{equation}
implying that $\Sigma_{ct}$ belongs to the cohomology of the linearized Slavnov-Taylor operator in the sector of the local integrated polynomials of dimension bounded by four. From the general results on the cohomology of Yang-Mills theories, see \cite{Piguet:1995er},  the counterterm $\Sigma_{ct}$ can be parametrized as follows 
\begin{equation}
	\Sigma_{ct} = a_{0}S_{\text{YM}}+a_{1}\frac{\lambda}{4!}(\phi^{a}\phi^{a})^{2}+a_{2}\frac{m_{\phi}^{2}}{2}\phi^{a}\phi^{a} + {\cal B}_{\Sigma}(\Delta^{-1})\;, \label{ctp}
\end{equation}
where $a_{0}, a_{1}, a_{2}$ are free arbitrary coefficients and  $\Delta^{-1}$ is an integrated polynomial in the fields and sources with dimension bounded by $4$ and with ghost number $-1$. The most general expression  for $\Delta^{-1}$  is given by 
\begin{eqnarray}
\Delta^{-1} &=& \int d^{4}x\; \left\{ a_{3}(\partial_{\mu}\bar{c}^{a} + K^{a}_{\mu})A^{a}_{\mu}
+ a_{4}L^{a}c^{a}
+ a_{5}\phi^{a}F^{a}
+ a_{6}\partial_{\mu}\varphi^{a}_{i}\partial_{\mu}\bar{\omega}^{a}_{i}
+ a_{7}\partial_{\mu}\eta^{aI}\partial_{\mu}\tilde{\theta}^{aI}
\right.
\nonumber \\
&&
\phantom{\int d^{4}x\;}
+ a_{8}\partial_{\mu}\bar{\omega}^{a}_{i}M^{a}_{\mu i}
+ a_{9}\bar{N}^{a}_{\mu i}\partial_{\mu}\varphi^{a}_{i}
+ a_{10}M^{a}_{\mu i}\bar{N}^{a}_{\mu i}
+ a_{11}V^{aI}\tilde{U}^{aI}
+ a_{12}m_{\phi}^{2}\varphi^{a}_{i}\bar{\omega}^{a}_{i}
\nonumber \\
&&
\phantom{\int d^{4}x\;}
+ a_{13}m_{\phi}^{2}\eta^{aI}\tilde{\theta}^{aI}
+ a_{14}gf^{abc}V^{aI}\phi^{b}\tilde{\theta}^{cI}
+ a_{15}gf^{abc}\tilde{U}^{aI}\phi^{b}\eta^{cI}
\nonumber \\
&&
\phantom{\int d^{4}x\;}
+ a_{16}gf^{abc}\partial_{\mu}A^{a}_{\mu}\varphi^{b}_{i}\bar{\omega}^{c}_{i}
+ a_{17}gf^{abc}A^{a}_{\mu}\partial_{\mu}\varphi^{b}_{i}\bar{\omega}^{c}_{i}
+ a_{18}gf^{abc}A^{a}_{\mu}\varphi^{b}_{i}\partial_{\mu}\bar{\omega}^{c}_{i}
\nonumber \\
&&
\phantom{\int d^{4}x\;}
+ a_{19}gf^{abc}A^{a}_{\mu}M^{b}_{\mu i}\bar{\omega}^{c}_{i}
+ a_{20}gf^{abc}A^{a}_{\mu}\bar{N}^{b}_{\mu i}\varphi^{c}_{i}
+ a_{21}gf^{abc}\partial_{\mu}A^{a}_{\mu}\eta^{bI}\tilde{\theta}^{cI}
\nonumber \\
&&
\phantom{\int d^{4}x\;}
+ a_{22}gf^{abc}A^{a}_{\mu}\partial_{\mu}\eta^{bI}\tilde{\theta}^{cI}
+ a_{23}gf^{abc}A^{a}_{\mu}\eta^{bI}\partial_{\mu}\tilde{\theta}^{cI}
\nonumber \\
&&
\phantom{\int d^{4}x\;}
+\mathbb{C}^{abcd}_{1}\phi^{a}\phi^{b}\varphi^{c}_{i}\bar{\omega}^{d}_{i}
+ \mathbb{C}^{abcd}_2\phi^{a}\phi^{b}\eta^{cI}\tilde{\theta}^{dI}
+ \mathbb{C}^{abcdIJKL}_3\eta^{aI}\tilde{\theta}^{bJ}\theta^{cK}\tilde{\theta}^{dL}
\nonumber \\
&&
\phantom{\int d^{4}x\;}
+ \mathbb{C}^{abcdIJKL}_4\eta^{aI}\tilde{\theta}^{bJ}\eta^{cK}\tilde{\eta}^{dL}
+ \mathbb{C}^{abcd}_5\varphi^{a}_{i}\bar{\varphi}^{b}_{i}\eta^{cI}\tilde{\theta}^{dI}
+ \mathbb{C}^{abcd}_6\omega^{a}_{i}\bar{\omega}^{b}_{i}\eta^{cI}\tilde{\theta}^{dI}
\nonumber \\
&&
\phantom{\int d^{4}x\;}
+ \mathbb{C}^{abcd}_7\varphi^{a}_{i}\bar{\omega}^{b}_{i}\theta^{cI}\tilde{\theta}^{dI}
+\mathbb{C}^{abcd}_8\varphi^{a}_{i}\bar{\omega}^{b}_{i}\eta^{cI}\tilde{\eta}^{dI}
+ \mathbb{C}^{abcdijkl}_9\varphi^{a}_{i}\bar{\omega}^{b}_{j}\varphi^{c}_{k}\bar{\varphi}^{d}_{l}
\nonumber \\
&&
\phantom{\int d^{4}x\;}
+ \left. \mathbb{C}^{abcdijkl}_{10}\varphi^{a}_{i}\bar{\omega}^{b}_{j}\omega^{c}_{k}\bar{\omega}^{d}_{l}
\right\}\;,
\end{eqnarray}
where $\left( \mathbb{C}^{abcd}_{1}, \mathbb{C}^{abcd}_2, \mathbb{C}^{abcdIJKL}_3, \mathbb{C}^{abcdIJKL}_4, \mathbb{C}^{abcd}_5, \mathbb{C}^{abcd}_6, \mathbb{C}^{abcd}_7, \mathbb{C}^{abcd}_8,  \mathbb{C}^{abcdijkl}_9, \mathbb{C}^{abcdijkl}_{10} \right)$ are arbitrary coefficients. After imposition of all other Ward identities it turns out that  the non-vanishing parameters which remain at the end of a lengthy algebraic analysis are:  
\begin{equation}
a_{3} = a_{6} = a_{7} = a_{8} = a_{9} = a_{10} = a_{17} = -a_{18} = a_{19} = a_{22} \neq 0   \;, \label{p1}
\end{equation}
as well as 
\begin{equation}
-a_{5} = a_{16} = a_{17}\neq 0 \;,  \qquad  \qquad a_{11}\neq 0 \;.   \label{p2}   
\end{equation}
Therefore, for the final expression of the invariant  counterterm one finds
\begin{eqnarray}
\label{fullct}
\Sigma_{ct} &=& \int d^{4}x \left\{ a_{0} F^{a}_{\mu\nu}F^{a}_{\mu\nu}
+ a_{1}\frac{\lambda}{4!}(\phi^{a}\phi^{a})^{2}
+ a_{2}\frac{m_{\phi}^{2}}{2}\phi^{a}\phi^{a}
+ a_{3}\left[ \frac{\delta S_{YM}}{\delta A^{a}_{\mu}}A^{a}_{\mu} 
+ \partial_{\mu}\bar{c}^{a}\partial_{\mu}c^{a}
\right.
\right.
\nonumber \\
&&
\phantom{\int d^{4}x}
+ K^{a}_{\mu}\partial_{\mu}c^{a}
-\bar{\varphi}^{a}_{i}\partial^{2}\varphi^{a}_{i}
+ \bar{\omega}^{a}_{i}\partial^{2}\omega^{a}_{i}
-\tilde{\eta}^{aI}\partial^{2}\eta^{aI}
+ \tilde{\theta}^{aI}\partial^{2}\theta^{aI}
- \bar{\varphi}^{a}_{i}\partial_{\mu}M^{a}_{\mu i}
\nonumber \\
&&
\phantom{\int d^{4}x}
+ N^{a}_{\mu i}\partial_{\mu}\bar{\omega}^{a}_{i}
+ \bar{M}^{a}_{\mu i}\partial_{\mu}\varphi^{a}_{i}
- \omega^{a}_{i}\partial_{\mu}\bar{N}^{a}_{\mu i}
- \bar{N}^{a}_{\mu i}N^{a}_{\mu i}
+ \bar{M}^{a}_{\mu i}M^{a}_{\mu i}
\nonumber \\
&&
\phantom{\int d^{4}x}
\left.
+ gf^{abc}\left(
- \partial_{\mu}c^{a}\varphi^{b}_{i}\partial_{\mu}\bar{\omega}^{c}_{i}
- \partial_{\mu}c^{a}\bar{N}^{b}_{\mu i}\varphi^{c}_{i}
+ \partial_{\mu}c^{a}M^{a}_{\mu i}\bar{\omega}^{c}_{i}
- \partial_{\mu}c^{a}\eta^{bI}\partial_{\mu}\tilde{\theta}^{cI}
\right)
\right]
\nonumber \\
&&
\phantom{\int d^{4}x}
+ a_{5}\left[gf^{abc}F^{a}\phi^{b}c^{c}
+ D^{ab}_{\mu}\phi^{b}D^{ac}_{\mu}\phi^{c}
+ m_{\phi}^{2}\phi^{a}\phi^{a}
+ \frac{\lambda}{3!}(\phi^{a}\phi^{a})^{2}
\right]
\nonumber \\
&&
\phantom{\int d^{4}x}
\left.
+ a_{11}\left(\tilde{V}^{aI}V^{aI} - \tilde{U}^{aI}U^{aI}\right)
\right\}\;.
\end{eqnarray}
It remains now to check that the counter term $\Sigma_{ct}$ can be reabsorbed into the initial action $\Sigma$, through a redefinition of the fields, sources and parameters, according to 
\begin{equation}
\label{renorm}
\Sigma(F,S,\xi)+\varepsilon \Sigma_{ct}(F,S,\xi) = \Sigma(F_{0},S_{0},\xi_{0}) + {\cal O}(\varepsilon^{2})\;,
\end{equation}
with
\begin{equation}
F_{0} = Z^{1/2}_{F}F\;, \qquad S_{0} = Z_{S}S \qquad \text{and} \qquad \xi_{0} = Z_{\xi}\xi\;, 
\end{equation}
where $\{F\}$ stands for all fields, $\{S\}$ for all sources and $\{xi\}$ for all parameters, {\it i.e.} $\xi=g,m_{\phi}, \lambda, \rho$. \\\\Therefore, by direct application of \eqref{renorm} we get
\begin{eqnarray}
&&
Z^{1/2}_{A} = 1+ \varepsilon\left(\frac{a_{0}}{2} + a_{3}\right) \\
&&
Z^{1/2}_{\phi}= 1 +\varepsilon a_{5}\\
&&
Z^{1/2}_{b} = Z^{-1/2}_{A} \\
&&
Z^{1/2}_{\bar{c}} =Z^{1/2}_{c} =Z^{-1/2}_{g} Z^{-1/4}_{A} \\
&&
Z^{1/2}_{\bar{\varphi}} =Z^{1/2}_{\varphi} =Z^{-1/2}_{g} Z^{-1/4}_{A}\\
&&
Z^{1/2}_{\bar{\omega}} =Z^{-1}_{g} \\
&&
Z^{1/2}_{\omega} =Z^{-1/2}_{A}\\
&&
Z^{1/2}_{\theta} = Z^{-1/2}_{A}\\
&&
Z^{1/2}_{\bar{\theta}} = Z^{-1}_{g}\\
&&
Z^{1/2}_{\eta} = Z^{1/2}_{\bar{\eta}} = Z^{-1/2}_{g}Z^{-1/4}_{A}\\
&&
Z_{N} =Z^{-1/2}_{A} \\
&&
Z^{1/2}_{\bar{N}} =Z^{-1}_{g} \\
&&
Z_{M} =Z_{\bar{M}} =Z^{-1/2}_{g}Z^{-1/4}_{A} \\
&&
Z_{V} = Z_{\bar{V}} = Z^{-1/2}_{\phi} Z^{1/2}_{g}Z^{1/4}_{A}\\
&&
Z_{U} = Z^{-1/2}_{\phi} \\
&&
Z_{\bar{U}} = Z^{-1}_{g}Z^{1/2}_{A}Z^{-1/2}_{\phi} \\
&&
Z_{K} =Z^{1/2}_{\bar{c}}\\
&&
Z_{F} = Z^{-1}_{\phi}Z^{1/4}_{A}Z^{-1/2}_{g}\;.
\end{eqnarray}
and
\begin{eqnarray}
&&
Z_{g} = 1-\varepsilon\frac{a_{0}}{2}\\
&&
Z_{m_{\phi}} = 1+\varepsilon a_{2}\\
&&
Z_{\lambda} = 1+\varepsilon a_{1}\\
&&
Z_{\rho} = (1+\varepsilon a_{11})Z^{-1}_{g}Z^{1/2}_{A}Z^{-1}_{\phi}\;.
\end{eqnarray}
These equations show that the invariant  counterterm  $\Sigma_{ct}$, eq.\eqref{fullct}, can be reabsorbed into the initial action $\Sigma$ through a multiplecative redefinition of the fields, sources and parameters. This concludes the algebraic proof of the all order renormalizability of $\Sigma$.

\end{appendix}


\begin{thebibliography}{9}

 \bibitem{Gribov:1977wm} 
  V.~N.~Gribov,
Nucl.\ Phys.\ B \textbf{139}, 1 (1978). 


\bibitem{Sobreiro:2005ec} 
  R.~F.~Sobreiro and S.~P.~Sorella,
  hep-th/0504095.
  
  
\bibitem{Vandersickel:2012tz} 
  N.~Vandersickel and D.~Zwanziger,
  Phys.\ Rept.\  {\bf 520}, 175 (2012)
  [arXiv:1202.1491 [hep-th]].


\bibitem{Singer:1978dk} 
  I.~M.~Singer,
  Commun.\ Math.\ Phys.\  {\bf 60}, 7 (1978).
  
  \cite{Dell'Antonio:1989jn,Dell'Antonio:1991xt,vanBaal:1991zw}
  
  
\bibitem{Dell'Antonio:1989jn} 
  G.~Dell'Antonio and D.~Zwanziger,
  Nucl.\ Phys.\ B {\bf 326}, 333 (1989).
  
\bibitem{Dell'Antonio:1991xt} 
  G.~Dell'Antonio and D.~Zwanziger,
  Commun.\ Math.\ Phys.\  {\bf 138}, 291 (1991).
  
\bibitem{vanBaal:1991zw} 
  P.~van Baal,
  Nucl.\ Phys.\ B {\bf 369}, 259 (1992).
  
  
     
  
\bibitem{Cucchieri:2007md}
  A.~Cucchieri and T.~Mendes,
  PoS LAT {\bf 2007} (2007) 297
  [arXiv:0710.0412 [hep-lat]].

\bibitem{Cucchieri:2007rg} 
  A.~Cucchieri and T.~Mendes,
  Phys.\ Rev.\ Lett.\  {\bf 100}, 241601 (2008)
  [arXiv:0712.3517 [hep-lat]].

\bibitem{Cucchieri:2008fc} 
  A.~Cucchieri and T.~Mendes,
  Phys.\ Rev.\ D {\bf 78}, 094503 (2008)
  [arXiv:0804.2371 [hep-lat]].
  
  
\bibitem{Cucchieri:2011ig} 
  A.~Cucchieri, D.~Dudal, T.~Mendes and N.~Vandersickel,
  Phys.\ Rev.\ D {\bf 85}, 094513 (2012)
  [arXiv:1111.2327 [hep-lat]].
  
\bibitem{Cucchieri:2012cb} 
  A.~Cucchieri, D.~Dudal and N.~Vandersickel,
  Phys.\ Rev.\ D {\bf 85}, 085025 (2012)
  [arXiv:1202.1912 [hep-th]].



\bibitem{Oliveira:2012eh} 
  O.~Oliveira and P.~J.~Silva,
  Phys.\ Rev.\ D {\bf 86}, 114513 (2012)
  [arXiv:1207.3029 [hep-lat]].
  
\bibitem{Oliveira:2008uf} 
  O.~Oliveira and P.~J.~Silva,
  Phys.\ Rev.\ D {\bf 79}, 031501 (2009)
  [arXiv:0809.0258 [hep-lat]].
  
\bibitem{Bornyakov:2009ug}
  V.~G.~Bornyakov, V.~K.~Mitrjushkin and M.~Muller-Preussker,
  Phys.\ Rev.\ D {\bf 81} (2010) 054503
  [arXiv:0912.4475 [hep-lat]].
  
\bibitem{Ilgenfritz:2010gu} 
  E.~-M.~Ilgenfritz, C.~Menz, M.~Muller-Preussker, A.~Schiller and A.~Sternbeck,
  Phys.\ Rev.\ D {\bf 83}, 054506 (2011)
  [arXiv:1010.5120 [hep-lat]].


\bibitem{Zwanziger:1988jt} 
  D.~Zwanziger,
  Nucl.\ Phys.\ B {\bf 321}, 591 (1989).
  
\bibitem{Zwanziger:1989mf} 
  D.~Zwanziger,
  Nucl.\ Phys.\ B {\bf 323}, 513 (1989).
  
\bibitem{Zwanziger:1992qr} 
  D.~Zwanziger,
  Nucl.\ Phys.\ B {\bf 399}, 477 (1993).
  


\bibitem{Dudal:2007cw} 
  D.~Dudal, S.~P.~Sorella, N.~Vandersickel and H.~Verschelde,
  Phys.\ Rev.\ D {\bf 77}, 071501 (2008)
  [arXiv:0711.4496 [hep-th]].
  
\bibitem{Dudal:2008sp} 
  D.~Dudal, J.~A.~Gracey, S.~P.~Sorella, N.~Vandersickel and H.~Verschelde,
  Phys.\ Rev.\ D {\bf 78}, 065047 (2008)
  [arXiv:0806.4348 [hep-th]].
  
\bibitem{Dudal:2011gd} 
  D.~Dudal, S.~P.~Sorella and N.~Vandersickel,
  Phys.\ Rev.\ D {\bf 84}, 065039 (2011)
  [arXiv:1105.3371 [hep-th]].
  
  
\bibitem{Gracey:2010cg} 
  J.~A.~Gracey,
  Phys.\ Rev.\ D {\bf 82}, 085032 (2010)
  [arXiv:1009.3889 [hep-th]].
  
\bibitem{Thelan:2014mza} 
  D.~J.~Thelan and J.~A.~Gracey,
  Phys.\ Rev.\ D {\bf 89}, 107701 (2014)
  [arXiv:1404.6364 [hep-th]].
  
  
\bibitem{Dudal:2010cd} 
  D.~Dudal, M.~S.~Guimaraes and S.~P.~Sorella,
  Phys.\ Rev.\ Lett.\  {\bf 106}, 062003 (2011)
  [arXiv:1010.3638 [hep-th]].
  
\bibitem{Dudal:2013wja} 
  D.~Dudal, M.~S.~Guimaraes and S.~P.~Sorella,
  arXiv:1310.2016 [hep-ph].
  
\bibitem{Mathieu:2008me} 
  V.~Mathieu, N.~Kochelev and V.~Vento,
  Int.\ J.\ Mod.\ Phys.\ E {\bf 18}, 1 (2009)
  [arXiv:0810.4453 [hep-ph]].
  
\bibitem{Canfora:2013zna} 
  F.~Canfora and L.~Rosa,
  Phys.\ Rev.\ D {\bf 88}, 045025 (2013)
  [arXiv:1308.1582 [hep-th]].
  
   
\bibitem{Fukushima:2013xsa} 
  K.~Fukushima and N.~Su,
  Phys.\ Rev.\ D {\bf 88}, 076008 (2013)
  [arXiv:1304.8004 [hep-ph]].
  
\bibitem{Canfora:2013kma} 
  F.~Canfora, P.~Pais and P.~Salgado-Rebolledo,
  arXiv:1311.7074 [hep-th].
  
\bibitem{Capri:2012ah} 
  M.~A.~L.~Capri, D.~Dudal, A.~J.~Gomez, M.~S.~Guimaraes, I.~F.~Justo, S.~P.~Sorella and D.~Vercauteren,
  Phys.\ Rev.\ D {\bf 88}, 085022 (2013)
  [arXiv:1212.1003 [hep-th]].
  
  
\bibitem{Capri:2013oja} 
  M.~A.~L.~Capri, D.~Dudal, M.~S.~Guimaraes, I.~F.~Justo, S.~P.~Sorella and D.~Vercauteren,
  Annals Phys.\ C {\bf 343}, 72 (2014)
  [arXiv:1309.1402 [hep-th]].
  
\bibitem{Fradkin:1978dv}
E.~H.~Fradkin and S.~H.~Shenker,
Phys.\ Rev.\ D {\bf 19}, 3682 (1979).



\bibitem{Capri:2014xea} 
  M.~A.~L.~Capri, D.~R.~Granado, M.~S.~Guimaraes, I.~F.~Justo, L.~F.~Palhares, S.~P.~Sorella and D.~Vercauteren,
  arXiv:1404.2573 [hep-th].

\bibitem{Capri:2014tta} 
  M.~A.~L.~Capri, M.~S.~Guimaraes, I.~F.~Justo, L.~F.~Palhares and S.~P.~Sorella,
  arXiv:1404.7163 [hep-th].
  
\bibitem{Baulieu:2008fy} 
  L.~Baulieu and S.~P.~Sorella,
  Phys.\ Lett.\ B {\bf 671}, 481 (2009)
  [arXiv:0808.1356 [hep-th]].
  
\bibitem{Dudal:2009xh} 
  D.~Dudal, S.~P.~Sorella, N.~Vandersickel and H.~Verschelde,
  Phys.\ Rev.\ D {\bf 79}, 121701 (2009)
  [arXiv:0904.0641 [hep-th]].
  
\bibitem{Sorella:2009vt} 
  S.~P.~Sorella,
  Phys.\ Rev.\ D {\bf 80}, 025013 (2009)
  [arXiv:0905.1010 [hep-th]].
  
\bibitem{Sorella:2010it} 
  S.~P.~Sorella,
  J.\ Phys.\ A {\bf 44}, 135403 (2011)
  [arXiv:1006.4500 [hep-th]].
  
\bibitem{Capri:2010hb} 
  M.~A.~L.~Capri, A.~J.~Gomez, M.~S.~Guimaraes, V.~E.~R.~Lemes, S.~P.~Sorella and D.~G.~Tedesco,
  Phys.\ Rev.\ D {\bf 82}, 105019 (2010)
  [arXiv:1009.4135 [hep-th]].
 
\bibitem{Dudal:2012sb} 
  D.~Dudal and S.~P.~Sorella,
  Phys.\ Rev.\ D {\bf 86}, 045005 (2012)
  [arXiv:1205.3934 [hep-th]].
  
\bibitem{Reshetnyak:2013bga} 
  A.~Reshetnyak,
  arXiv:1312.2092 [hep-th].
  
\bibitem{Baulieu:2009ha} 
  L.~Baulieu, D.~Dudal, M.~S.~Guimaraes, M.~Q.~Huber, S.~P.~Sorella, N.~Vandersickel and D.~Zwanziger,
  Phys.\ Rev.\ D {\bf 82}, 025021 (2010)
  [arXiv:0912.5153 [hep-th]].
  
\bibitem{Cucchieri:2014via} 
  A.~Cucchieri, D.~Dudal, T.~Mendes and N.~Vandersickel,
  arXiv:1405.1547 [hep-lat].
  
\bibitem{Zwanziger:2010iz} 
  D.~Zwanziger,
  Phys.\ Rev.\ D {\bf 81}, 125027 (2010)
  [arXiv:1003.1080 [hep-ph]].
  
  
 
  
\bibitem{Maas:2011yx} 
  A.~Maas,
  PoS FACESQCD {\bf }, 033 (2010)
  [arXiv:1102.0901 [hep-lat]].
  
  
  
\bibitem{Maas:2010nc} 
  A.~Maas,
  Eur.\ Phys.\ J.\ C {\bf 71}, 1548 (2011)
  [arXiv:1007.0729 [hep-lat]].
  
  
    
\bibitem{Furui:2006ks} 
  S.~Furui and H.~Nakajima,
  Phys.\ Rev.\ D {\bf 73}, 074503 (2006).
  
\bibitem{Parappilly:2005ei} 
  M.~B.~Parappilly, P.~O.~Bowman, U.~M.~Heller, D.~B.~Leinweber, A.~G.~Williams and J.~BZhang,
  Phys.\ Rev.\ D {\bf 73}, 054504 (2006)
  [hep-lat/0511007].
  
  
  \bibitem{axel}
  A.~Maas, {\it Private Communication}.
  
  \bibitem{Orlando}
  O.~Oliveira, {\it Private Communication}.
  
\bibitem{Dudal:2013vha}
  D.~Dudal, M.~S.~Guimaraes, L.~F.~Palhares and S.~P.~Sorella,
  arXiv:1303.7134 [hep-ph].
  
\bibitem{Pelaez:2014mxa}
  M.~Pel‡ez, M.~Tissier and N.~Wschebor,
  arXiv:1407.2005 [hep-th].
  
  \bibitem{MAG-wip}
  M.~A.~L.~Capri {\it et al}, {\it Work in progress}.
  
\bibitem{Baulieu:2009xr} 
  L.~Baulieu, M.~A.~L.~Capri, A.~J.~Gomez, V.~E.~R.~Lemes, R.~F.~Sobreiro and S.~P.~Sorella,
  Eur.\ Phys.\ J.\ C {\bf 66}, 451 (2010)
  [arXiv:0901.3158 [hep-th]].



\bibitem{Piguet:1995er} 
  O.~Piguet and S.~P.~Sorella,
  Lect.\ Notes Phys.\ M {\bf 28}, 1 (1995).


\end{thebibliography}
\end{document}